\documentclass[floats,aps,showpacs,twocolumn]{revtex4-1}

\usepackage[dvips]{graphicx}
\usepackage{amsfonts}
\usepackage{amsmath,amssymb}
\usepackage{dsfont}
\usepackage{placeins}
\usepackage{afterpage}
\usepackage{flafter}

\begin{document}

\newcommand{\iexp}{\text{i}}
\newcommand{\diff}{\text{d}}
\newcommand{\heff}{\hbar}
\newcommand{\hw}{\hbar\omega}
\newcommand{\Rac}{ R^{\text{ac}} }
\newcommand{\Ereg}{E^{\text{reg}}}
\newcommand{\Nreg}{N_{\text{reg}}}
\newcommand{\mmax}{m_{\text{max}}}
\newcommand{\pch}{\bar{p}_{\text{ch}}}
\newcommand{\betaeff}{\beta_{\text{eff}}}
\newcommand{\betafit}{\beta_{\text{fit}}}

\title{Statistical mechanics of Floquet systems with regular and chaotic states}

\author{Roland Ketzmerick}
\author{Waltraut Wustmann}
\affiliation{Institut f\"ur Theoretische Physik,
             Technische Universit\"at Dresden, 01062 Dresden, Germany}
\date{\today}

\begin{abstract}
We investigate the asymptotic state of
time-periodic quantum systems with regular and chaotic
Floquet states weakly coupled to a heat bath.
The asymptotic occupation probabilities of these two types of states
follow fundamentally different distributions.
Among regular states the probability decreases from the
state in the center of a regular island to the outermost state by 
orders of magnitude,
while chaotic states have almost equal probabilities.
We derive an analytical expression for the occupations of regular states of kicked
systems,
which depends on the winding numbers of the regular tori 
and the parameters temperature and driving frequency.
For a constant winding number within a regular island 
it simplifies to  Boltzmann-like weights $\exp(-\betaeff \Ereg_m)$, 
similar to time-independent systems.
For this we introduce the regular energies $\Ereg_m$ of the quantizing tori and 
an effective winding-number-dependent temperature $1/\betaeff$, 
different from the actual bath temperature.
Furthermore, the occupations of other typical Floquet states in a mixed phase 
space are studied, i.e.~regular states on nonlinear resonances, beach states,
and hierarchical states, giving rise to distinct features in the occupation
distribution.
Avoided crossings involving a regular state lead to 
drastic consequences for the entire set of occupations.
We introduce a simplified rate model whose analytical solutions describe the
occupations quite accurately.
\end{abstract}

\maketitle

\section{Introduction}

The response of a dynamical system to a time-periodic driving force
is ubiquitous in both classical and quantum mechanics
and plays a fundamental role in many physical and technical applications.
It opened the field
for the coherent control of atoms and molecules~\cite{RicZha2000},
the optimal control of chemical reactions~\cite{BruSha2003},
or the manipulation of semiconductor-nanodevices and heterostructures
in solids~\cite{Phi1994}.
Under realistic, nonidealized conditions real physical systems interact
with their environment.
If the environment contains a vast number of degrees of freedom,
the full dynamics of the composite system is not traceable.
The system is then interpreted as an open subsystem in mutual contact
with a heat bath and the dynamics of the subsystem is characterized
by its reduced density operator $\rho$.
To evaluate the evolution of $\rho$
for an open quantum system in a time-varying, strong external field
is a nontrivial task, as it is permanently driven out of equilibrium.
For only very few systems exact analytical solutions of the
damped dynamics are feasible, in particular a
driven two-level system~\cite{GriSasStoWei1993_GriSasHaeWei1995}
and a driven harmonic oscillator~\cite{GraHue1994, ZerHae1995}.
In general systems it is studied numerically,
e.g.\ with focus on tunneling, see Ref.~\cite{GriHae1998} and references therein.
Especially in the regime of weak interaction with the environment,
standard methods, originally established for time-independent
quantum systems, have been adapted to the demands of time-periodic
systems~\cite{BluETAL1991,KohDitHae1997,BreHubPet2000,Koh2001,HonKetKoh2009}.

The final state of the relaxation process has so far not received as much
attention as transient phenomena,
although this can be ranked as even more fundamental and is in fact
a core question of statistical mechanics.
The usual thermodynamic concepts for the equilibrium state of time-independent
systems are not applicable, such as the canonical distribution of Boltzmann
weights, reached in the stationary limit of a time-independent system
that is weakly coupled to a heat bath.
The Boltzmann weights $e^{-\beta E_n}$ of the eigenstates are unique functions
of the eigenenergy with the temperature $1/\beta = k_B T$ of the heat bath
as the only relevant parameter,
whereas microscopic details of the weak coupling play no role.
Such a stationary limit, in the sense of convergence to time-independent
values for all dynamical variables, is not encountered in a periodically
driven system, where energy is permanently exchanged between the driven system
and the environment.
Instead, the relaxation process finally leads to
an asymptotic state that adopts the periodicity of the driving, and
that in general depends on the microscopic details of the coupling.
The density operator of the time-periodic subsystem is best represented
in the Floquet state basis.
The Floquet states are quasi-periodic solutions of the Schr\"odinger equation
for the time-periodic Hamiltonian without the coupling to the environment.
In the Floquet basis the evolution equation for the density matrix
can be approximated within the Floquet-Markov
approach~\cite{BluETAL1991,KohDitHae1997,BreHubPet2000,Koh2001,HonKetKoh2009}
by a Markovian quantum master equation.
In the long-time limit of the evolution the Floquet states are
populated with asymptotic occupation probabilities that can be determined
from a system of rate equations.
Beyond the numerical evaluation of such a master equation,
an intuitive understanding of these Floquet occupations is still lacking.

A related problem occurs at avoided crossings,
which are ubiquitous in the quasienergy spectra of generic Floquet systems,
since the quasienergies are bounded within a finite interval,
$0 \leq \varepsilon < \hbar\omega$, where $\omega = 2\pi/\tau$ is the driving
frequency and $\tau$ is the driving period.
As a consequence, the number of avoided crossings grows without limit for
increasing Hilbert-space dimension, leading to a breakdown of the adiabatic
theorem~\cite{HonKetKoh1997}.
In the presence of a heat bath this problem is approached
in Ref.~\cite{HonKetKoh2009} where it is shown that the reduced density
operator $\rho$ is not affected by a small avoided crossing, provided that it
is smaller than a specific effective coupling parameter
and so is not `resolved' by the heat bath.
These findings justify the unevitable truncation of the,
in general, infinite Hilbert space dimension in numerical implementations.

One way to tackle the general challenge of finding the Floquet occupations
beyond their numerical evaluation is to study the semiclassical
regime of one-dimensional driven systems.
In their classical limit regular and chaotic motions generically coexist.
This is most clearly reflected in phase space, where regular trajectories
evolve on invariant tori and chaotic trajectories fill the remaining
phase-space regions.
According to the semiclassical eigenfunction
hypothesis~\cite{Per1973_Ber1977_Vor1979} almost all Floquet states can be
classified as either regular or chaotic,
provided that the phase-space regions are larger than the Planck constant.
The regular states localize on the regular tori
and the chaotic states typically spread out over the chaotic region.
For a driven particle in a box coupled to a heat bath
the Floquet occupations of regular and chaotic states were found to follow
different statistical distributions~\cite{BreHubPet2000}.
The regular states, which in this example differ only slightly from the
eigenstates of the undriven system, carry almost Boltzmann weights, whereas
all chaotic states have nearly the same occupation probability.

In this paper we concentrate on situations characteristic for strong driving,
where both phase space and Floquet states are strongly perturbed compared
to the originally time-independent system.
We demonstrate that the Floquet occupations of the states in a regular
island under these conditions deviate considerably from the Boltzmann result.
For kicked systems, making use of some reasonable assumptions, we derive an
analytical expression for the regular occupations.
In many cases it can be well approximated by
weights of the Boltzmann type $e^{-\betaeff \Ereg_m}$.
This requires the introduction of the regular energies $\Ereg_m$,
which are semiclassical invariants of the quantizing tori of the regular
island, and the parameter $1/\betaeff$, which is an effective temperature
depending on the winding number of the regular island.
Furthermore, we give an overview and interpretation for the occupations of
other typical Floquet states in a mixed phase space, such as states on a
resonance island chain, beach states, and hierarchical states.

Avoided crossings in the Floquet spectrum can lead to severe changes
in the occupations if they are larger than an effective coupling parameter.
The effect can be intuitively explained
by a set of effective rate equations with an additional rate $\Rac$
between the states of the avoided crossing~\cite{HonKetKoh2009}.
It can be exploited for a switching mechanism in driven quantum systems,
e.g.\ a weakly driven bistable system~\cite{KetWus2009}.
For the above occupation distributions for regular and chaotic states we
demonstrate drastic consequences if a regular state has an avoided crossing
with either a regular or a chaotic state.
We introduce a simplified rate model whose analytical solution describes
the Floquet occupations accurately.

The paper is organized as follows:
in Sec.~\ref{sec:FloquetMarkov} the microscopic model of driven dissipative
systems and the Floquet-Markov description of its asymptotic state are sketched
and the relevant coupling operators are introduced.
Section~\ref{sec:Occupations} presents general occupation characteristics
for the example of a driven quartic oscillator (Sec.~\ref{sec:Driven})
and a kicked rotor (Sec.~\ref{sec:Kicked}). They are related to the
corresponding rate matrices (Sec.~\ref{sec:Rates}).
In Sec.~\ref{sec:RegularStates} we derive an analytical expression for the
regular occupations of kicked systems,
depending on the winding numbers of the regular tori.
For a constant winding number a simplification
to the Boltzmann-like weights $e^{-\betaeff \Ereg_m}$ is shown
(Sec.~\ref{sec:approx2_betaeff}).
An example where this is not possible is also discussed
(Sec.~\ref{sec:examples}).
Section~\ref{sec:AddStructs} gives an overview of the occupation
characteristics of other types of Floquet states.
In Sec.~\ref{sec:AC} the influence of avoided crossings on the Floquet
occupations is demonstrated, which are compared to the analytical solutions of
a simplified rate model (Sec.~\ref{sec:AC_model}).
Section~\ref{sec:Summary} summarizes the paper.

\section{Master equation in time-periodic systems}\label{sec:FloquetMarkov}

The coupling of a quantum system with the Hamiltonian $H_s(t)$ to a heat bath
is modelled in a standard way by the composite Hamiltonian~\cite{Wei1999}
\begin{equation}\label{eq:Hamiltonian_tot}
H(t) = H_s(t) + H_b + H_{sb}
.
\end{equation}
Herein, the bath Hamiltonian
$H_b = \sum_n \left( \frac{p_n^2}{2m_n}
+ \frac{m_n \omega_n^2}{2}  x_n^2 \right)$
describes an ensemble of noninteracting harmonic oscillators
coupled via the interaction Hamiltonian $H_{sb}$ to the system.
In spatially extended systems this interaction is commonly assumed to be bilinear,
\begin{equation}\label{eq:H_sb}
H_{sb} = A \sum_n c_n x_n
,
\end{equation}
with some coupling operator $A$ of the system.
The properties of the system-bath coupling are specified by the spectral
density of the bath
$J(\omega) := \frac{\pi}{2} \sum_n \frac{c_n^2}{m_n \omega_n}
\left[ \delta(\omega-\omega_n) - \delta(\omega+\omega_n) \right]$.
In the continuum limit the spectral density is assumed to be a smooth
function which is linear for an Ohmic bath.
An exponential cutoff beyond the spectral mode $\omega_c$ leads to
$J(\omega) = \eta \omega \, e^{-\left|\omega\right|/\omega_c}$,
where $\eta$ is proportional to the classical damping coefficient.

In the absence of the heat bath the solutions of the time-dependent 
Schr\"odinger equation for the isolated system with the $\tau$-periodic
Hamiltonian
\begin{equation}
H_s(t+\tau) = H_s(t)
\end{equation}
are the Floquet states $|\psi_i(t)\rangle$.
These can be factorized into a product
\begin{equation}
|\psi_i(t)\rangle = e^{-\iexp \varepsilon_i t/\hbar} | u_i(t) \rangle
\end{equation}
of a phase factor with the quasienergy $\varepsilon_i$
and a periodic state vector $|u_i(t)\rangle$,
\begin{equation}
|u_i(t+\tau)\rangle = |u_i(t)\rangle
,
\end{equation}
with the period $\tau$ of the Hamiltonian.

In the presence of the heat bath the state of the system is described by
the reduced density operator $\rho(t)$.
Its equation of motion for time-periodic quantum systems 
has been derived within the Floquet-Markov
approach~\cite{BluETAL1991,KohDitHae1997,BreHubPet2000,Koh2001,HonKetKoh2009}:
herein the Floquet formalism ensures a non-perturbative treatment
of the coherent dynamics of the driven system. 
The density operator is represented in the set
of the time-periodic state vectors $|u_i(t)\rangle$,
which form a complete orthonormal basis at all times $t$.
The coupling to the heat bath is treated perturbatively in second order
of $H_{sb}$, which is valid in the limit of weak coupling between the
driven system and the bath.
This approximation requires a rapid decay of bath correlations compared
to the typical relaxation time of the system.
In this paper we use a cutoff frequency $\omega_c = 100 \omega$,
which is large compared to the frequency $\omega = 2\pi/\tau$ of the driving.
In the following we restrict the discussion to the limit of large times,
much larger than the relaxation time.
In this limit the density-matrix elements
$\rho_{ij} = \langle u_i(t)|\rho(t)|u_j(t)\rangle$
are approximated as time-independent~\cite{KohDitHae1997, HonKetKoh2009}.
Note that the corresponding density operator,
$\sum_{i,j} | u_i(t) \rangle \rho_{ij} \langle u_j(t) |$,
is still time-periodic because of the inherent time-dependence of the
$|u_i(t)\rangle$.
In this paper we restrict to the weak-coupling regime,
where the system-bath coupling is small compared to all quasienergy spacings
of a truncated Hilbert space (see discussion below).
The Floquet occupations $p_i \equiv \rho_{ii}$ then
obey the set of rate equations
\begin{equation}\label{eq:RGS_rho_diag}
0 = - p_{i} \sum_{j} R_{ij}  + \sum_j p_{j} R_{ji}
,
\end{equation}
which are independent of the damping coefficient $\eta$.
Note that the rate equations beyond this weak-coupling regime
would also contain the non-diagonal elements $\rho_{ij}$ ($i \neq j$).
The rates
\begin{equation}\label{eq:R_ik}
R_{ij} := \frac{\pi}{\hbar} \sum_K \left| A_{ij}(K) \right|^2
g(\varepsilon_j-\varepsilon_i - K \hw)
,
\end{equation}
that describe bath-induced transitions between the Floquet states,
use the Fourier coefficients
\begin{equation}\label{eq:x_ijK_def}
A_{ij}(K) = \frac{1}{\tau} \int_0^{\tau} \diff t e^{-\iexp \omega K t} A_{ij}(t)
\end{equation}
of the time-periodic matrix elements
\begin{equation}\label{eq:x_ijt_def}
A_{ij}(t) = \left\langle u_i(t) \right| A \left| u_j(t) \right\rangle
.
\end{equation}
The correlation function $g(E) = \pi^{-1} n_\beta(E) J(E/\hbar)$
of the bath coupling operator contains the spectral density $J(\omega)$
and the thermal occupation number $n_\beta(E)$ of the boson bath
with temperature $1/\beta$.

The reduction to the set of rate equations~\eqref{eq:RGS_rho_diag}
seems possible only for systems with a finite dimension of the Hilbert space,
since otherwise the quasienergies densely fill the interval $[0,\hw)$.
However, as demonstrated in Ref.~\cite{HonKetKoh2009},
near degeneracies much smaller than the coupling strength
are not resolved by the interaction to the heat bath
and do not influence the asymptotic density operator.
The Hilbert dimension can therefore be truncated, keeping only those
Floquet states of non-negligible occupation.

Equation~\eqref{eq:RGS_rho_diag} is formally identical to the familiar system
of rate equations describing the equilibrium state in time-independent systems.
In contrast, however, specifics of the time-periodic system are present in
the rates, Eq.~\eqref{eq:R_ik}, whose structure does in general not
allow a detailed balance relation~\cite{Koh2001}.

\subsection{Coupling operator for extended and for cyclic systems}

For extended systems we assume as usual~\cite{Wei1999}
the linear coupling operator
\begin{equation}
A = x
\end{equation}
in Eq.~\eqref{eq:H_sb} for the interaction Hamiltonian $H_{sb}$ 
with the heat bath.

For cyclic systems defined on the unit interval $[0,1)$ with periodic boundary
conditions in $x$ the coupling operator $x$ would be discontinuous at the
borders of the interval and the coupling to the heat bath would therefore
not be homogeneous.
An adapted coupling scheme with the interaction Hamiltonian
\begin{equation}\label{eq:H_sb_cyclic}
H_{sb} = \int_0^{2\pi} \diff \alpha \frac{\sqrt{2}}{2\pi} \sin(2\pi x + \alpha)
\sum_n c_n x_n \delta(\alpha - \alpha_n)
\end{equation}
has been proposed in Ref.~\cite{Coh1994} for such situations.
The angles $\alpha_n$ characterize the individual bath oscillators
and are equidistributed in the interval $[0,2\pi)$.
The new spectral density
\begin{equation}
J(\omega,\alpha) =
\frac{\pi}{2} \sum_n \frac{c_n^2}{m_n \omega_n}  \delta(\alpha-\alpha_n)
\left[ \delta(\omega-\omega_n) - \delta(\omega+\omega_n) \right]
\end{equation}
hence factorizes into independent factors,
$J(\omega, \alpha)= J(\omega)/(2\pi)$,
with the spectral density $J(\omega)$ as defined above and the homogeneous
angular density $1/(2\pi)$.
The spatially periodic interaction Hamiltonian~\eqref{eq:H_sb_cyclic}
is continuous and, by virtue of the equidistributed angles $\alpha_n$,
models the interaction with a homogeneous environment, where no position is
singled out.
Making use of the trigonometric addition theorem
$\sin(2\pi x + \alpha) = \sin(2\pi x)\cos(\alpha) + \cos(2\pi x)\sin(\alpha)$,
the interaction Hamiltonian~\eqref{eq:H_sb_cyclic} leads to the
same system of rate equations, Eq.~\eqref{eq:RGS_rho_diag}.
But now the rates
\begin{equation}\label{eq:R_ik_cyclic}
R_{ij} = R_{ij}^{(1)} + R_{ij}^{(2)}
\end{equation}
are composed of
two independent contributions from the simpler coupling operators
\begin{eqnarray}
\label{eq:A1_cyclic}
A^{(1)} &=& \sin(2\pi x)/(2\pi), \\
\label{eq:A2_cyclic}
A^{(2)} &=& \cos(2\pi x)/(2\pi),
\end{eqnarray}
respectively, to be used in the interaction Hamiltonian $H_{sb}$
in Eq.~\eqref{eq:H_sb}.
For this result we use that the mixed second-order term
$\int_0^{2\pi} \diff \alpha \int_0^{2\pi} \diff \alpha' \cos(\alpha) \sin(\alpha')
\delta(\alpha-\alpha_n) \delta(\alpha'-\alpha_n)
= \int_0^{2\pi} \diff \alpha \cos(\alpha) \sin(\alpha) = 0$
vanishes, while the other two terms give rise to the operators
in Eqs.~\eqref{eq:A1_cyclic} and~\eqref{eq:A2_cyclic}.

\section{Occupations of regular and chaotic Floquet states}%
\label{sec:Occupations}

In this section we study the Floquet occupations for two classes of periodically
driven systems, the additively driven quartic oscillator
as a representative of a continuously driven system
and the quantum kicked rotor from the class of kicked systems.
In both cases we demonstrate that the Floquet occupations of regular and
chaotic Floquet states follow tremendously different distributions.
Similar observations have been made for a driven quantum
particle in a box~\cite{BreHubPet2000}.
While there it was found that regular states carry Boltzmann weights,
we find typically significant deviations.
Numerical results will be presented in this section and the
quantitative analysis of the occupations of the regular states will
be deferred to Sec.~\ref{sec:RegularStates}.

\subsection{Driven Quartic Oscillator}\label{sec:Driven}

\begin{figure}[tb]
\includegraphics[width=8.5cm]{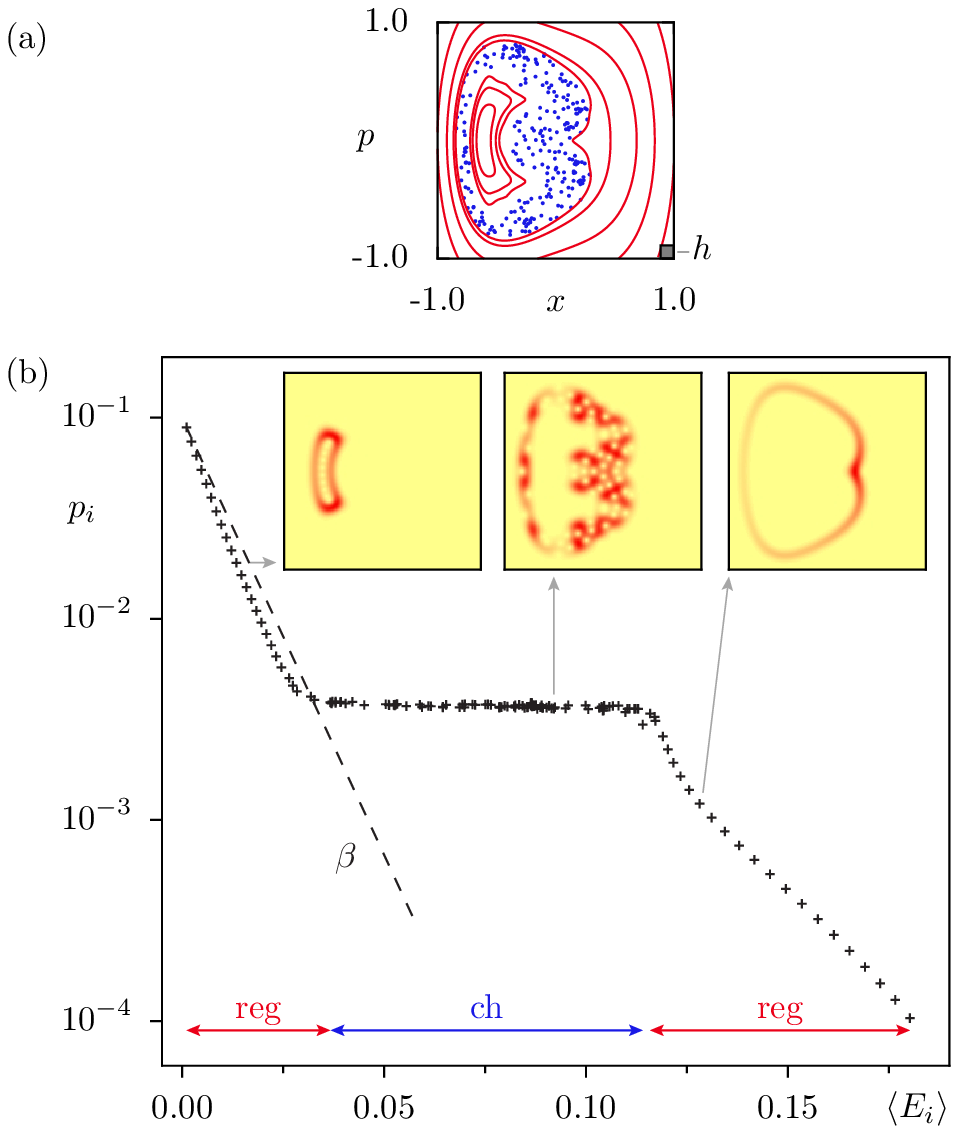}
\caption{
(Color online) 
(a) Stroboscopic Poincar\'{e} section of the classical phase space of
the driven osillator, Eq.~\eqref{eq:H_quart}.
The size of the chosen dimensionless Planck constant $h$ is indicated
in the lower right corner.
(b) Floquet occupations $p_i$ vs.\ cycle-averaged energies
$\langle E_i \rangle$ compared to the Boltzmann-like prediction
$\exp(-\beta \langle E_i \rangle)$ (dashed line).
The insets show Husimi representations of a regular Floquet state
localized in the central island, a chaotic Floquet state,
and a regular state on a surrounding torus.
The parameters are $\kappa = 0.2$, $\omega=5/6$, 
$\heff = 0.002$, and $\beta = 100$.
}
\label{fig:phSp_Emean_rho_Ereg_rho_quart}
\end{figure}

We consider as an example of a continuously driven system
the additively driven quartic oscillator with the Hamiltonian
\begin{equation}
H_s(t) =  \frac{p^2}{2m} + V_0
\left( \frac{x^4}{x_0^4}  + \kappa \frac{x}{x_0} \cos(\omega t) \right)
.
\end{equation}
We introduce the dimensionless quantities
$\tilde x = x / x_0$,
$\tilde H = H / V_0$,
$\tilde p = p / p_0$ with $p_0 = \sqrt{m V_0}$,
$\tilde t = t \cdot p_0 / (x_0 m)$,
$\tilde \omega = \omega \cdot (x_0 m) / p_0$,
and also $\tilde \hbar = \hbar / (x_0 p_0)$,
the ratio of the Planck constant to a typical phase-space area.
In the following we omit the overtilde and then the dimensionless Hamiltonian reads
\begin{equation}\label{eq:H_quart}
H_s(t) =  \frac{p^2}{2} + x^4  + \kappa x \cos( \omega t)
.
\end{equation}
At $\kappa=0.2$ and $\omega = 5/6$
the stroboscopic Poincar\'{e} section of the
phase space $(x,p)$ at integer
multiples $t=n\tau$ of the driving period, $\tau = 2\pi/\omega$,
features a chaotic domain, see Fig.~\ref{fig:phSp_Emean_rho_Ereg_rho_quart}(a).
Furthermore, there are two distinct regular regions:
first the highly excited tori, which are only slightly influenced by the
driving, and second a regular island embedded in the chaotic sea.

The Floquet states are determined using the
$(t,t')$-technique~\cite{PfeLev1983_PesMoi1993}.
Their energy expectation value, 
$\langle \psi_i(t) |H_s(t)| \psi_i(t)\rangle 
= \langle u_i(t) |H_s(t)| u_i(t)\rangle$,
oscillates with the period of the driving.
It is convenient to introduce the cycle-averaged energy
\begin{equation}\label{eq:Emean}
\langle E_i \rangle :=
\frac{1}{\tau} \int_t^{t+\tau} \diff t' \langle u_i(t') |H_s(t')| u_i(t')\rangle
- E_0
.
\end{equation}
An energy shift $E_0$ is
determined by the classical periodic orbit at the center of the regular island,
such that there the cycle-averaged energy is zero.

For a sufficiently small value of $h$
almost all Floquet states can be classified as either regular or chaotic
according to the semiclassical eigenfunction
hypothesis~\cite{Per1973_Ber1977_Vor1979}.
The regular states are localized on the regular island and
can be ordered by a quantum number $m$, whereas the chaotic states typically
spread over the entire chaotic phase-space area and fluctuate irregularly.
The different types of states are visualized in phase space
by means of their Husimi representation
$H_\psi(x,p) := \left| \left\langle \alpha(x,p) \right.
\left| \psi \right\rangle \right|$, i.e.\ the
projection onto the coherent states $| \alpha(x,p) \rangle$
centered at the phase space points $(x,p)$,
see insets in Fig.~\ref{fig:phSp_Emean_rho_Ereg_rho_quart}(b).
In the example of Fig.~\ref{fig:phSp_Emean_rho_Ereg_rho_quart}
the small central island 
is about $26$ times larger than the dimensionless Planck constant $h$,
indicated in the lower right corner of 
Fig.~\ref{fig:phSp_Emean_rho_Ereg_rho_quart}(a).
Thus the central island supports $26$ regular states.
Besides, there are $97$ chaotic states, spreading over the chaotic region.
It is surrounded by regular tori, on which a further group of infinitely many
regular states are localized.
The cycle-averaged energies $\langle E_i \rangle$ of these three different types
of Floquet states form distinct intervals with only small overlaps,
as indicated by the arrows in Fig.~\ref{fig:phSp_Emean_rho_Ereg_rho_quart}(b).
The regular states of the central island are lowest in cycle-averaged energy,
followed by the chaotic states and finally the high excited
regular states of the surrounding tori.

The oscillator coupled to a heat bath is treated as explained
in Sec.~\ref{sec:FloquetMarkov} with the linear coupling operator $A=x$
for this spatially extended system.
We restrict the consideration to the weak-coupling regime,
where the occupations are well described by the rate
equation~\eqref{eq:RGS_rho_diag}.
In Fig.~\ref{fig:phSp_Emean_rho_Ereg_rho_quart}(b) the resulting Floquet
occupations $p_i$ are shown as functions of the cycle-averaged energies 
$\langle E_i \rangle$.
The monotonously falling occupations at low values of
$\langle E_i \rangle$ belong to the central regular island,
with the state in the center of the island having the highest occupation.
At intermediate values of $\langle E_i \rangle$ one finds
the occupations of the chaotic states, that fluctuate around a mean
value $\pch$, with a very small variance compared
to the range of occupations of the regular states.
This is similar to the observation for chaotic states in
Ref.~\cite{BreHubPet2000} for a driven particle in a box.
At high values of $\langle E_i \rangle$, there are again monotonously falling
occupations belonging to the regular states of the surrounding tori.

The observed characteristics of the occupations $p_i$ are clearly in contrast
to the naive expectation
$p_i \sim e^{-\beta \left\langle E_i\right\rangle}$
motivated by the Boltzmann weights of equilibrium thermodynamics.
Note that even the occupations of the (low-energy) regular
states notably differ from the Boltzmann result, indicated by the
dashed line in Fig.~\ref{fig:phSp_Emean_rho_Ereg_rho_quart}(b).
A quantitative analysis of these observations will be presented in
Sec.~\ref{sec:RegularStates} for the numerically more convenient kicked rotor.

\subsection{Kicked rotor}\label{sec:Kicked}

\begin{figure}[tb]
\includegraphics[width=8.5cm]{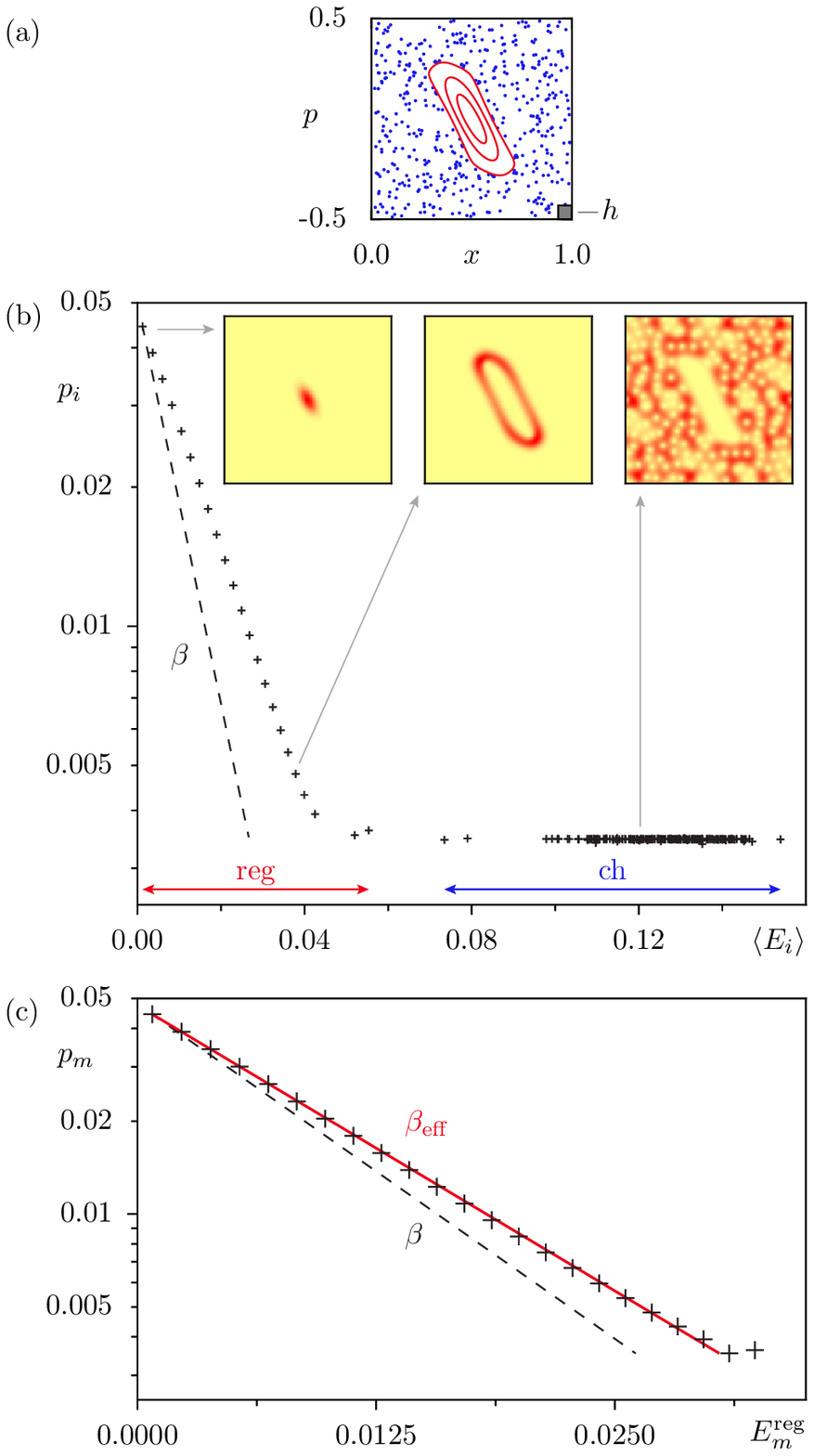}
\caption{
(Color online) 
(a) Stroboscopic Poincar\'{e} section of the classical phase space of
the kicked rotor, Eq.~\eqref{eq:H_kiRo}, for $\kappa = 2.9$.
The size of the chosen dimensionless Planck constant $h$ is indicated
in the lower right corner.
(b) Floquet occupations $p_i$ vs.\ cycle-averaged energies $\langle E_i \rangle$
compared to the Boltzmann-like prediction
$\exp(-\beta \langle E_i \rangle)$ (dashed line).
The insets show Husimi representations of two regular and a
chaotic Floquet state.
(c) Floquet occupations $p_m$ of regular states vs.\ regular energies $\Ereg_m$,
defined in Eq.~\eqref{eq:Ereg_m},
compared to the Boltzmann-like prediction Eq.~\eqref{eq:dist_preg}
with the inverse effective temperature $\betaeff = 0.85 \beta$
of Eq.~\eqref{eq:betaeff} (red solid line),
compared to the inverse bath temperature $\beta$ (dashed line).
The parameters are $h = 1/210$ and $\beta = 100$.
}
\label{fig:phSp_Emean_rho_Ereg_rho}
\end{figure}

Kicked quantum systems feature all essential
phase-space characteristics of periodically driven systems.
They allow for a simplified numerical and conceptual treatment.
As a paradigmatic model for a driven system with a mixed phase space
we consider here the quantum kicked rotor.
Its dynamics is generated by the Hamiltonian
\begin{equation}\label{eq:H_kicked}
H_s(t) = T(p) + V(x) \cdot \tau \sum_n \delta(t-n \tau)
,
\end{equation}
with the kinetic energy
$T(p) = p^2/(2m)$
and the potential
$V(x) = V_0 \cos(2\pi x/x_0)$
acting at the kicks.
We study it on a two-torus with dimensions $x_0$ and $p_0=m x_0 /\tau$.
We make the Hamiltonian dimensionless by a similar transformation as in
Sec.~\ref{sec:Driven},
where now $\tilde H = H \cdot m/p_0^2$
and the dimensionless kick period is $\tau = 1$.
With the rescaled kick strength $\kappa = V_0 \cdot (2\pi)^2 m/p_0^2$
the Hamiltonian reads
\begin{equation}\label{eq:H_kiRo}
 H_s(t) = \frac{p^2}{2} + \frac{\kappa}{(2\pi)^2} \cos(2\pi x) \sum_n \delta(t-n)
.
\end{equation}
In an intermediate regime of the kick strength
the Poincar\'{e} section of the phase space features
a regular island embedded in the chaotic sea,
see Fig.~\ref{fig:phSp_Emean_rho_Ereg_rho}(a) for $\kappa=2.9$.

The Floquet states are evaluated as eigenstates of the time evolution
operator $U$ over one period $\tau$, which factorizes into a potential and 
a kinetic part $U = e^{-\iexp \tau V(x)/\heff} e^{-\iexp \tau T(p)/\heff}$.
The quantization on the two-torus relates the effective Planck constant $h$ to 
the dimension $N$ of the Hilbert space by the condition $h = 2\pi\heff = 1/N$.
For $h=1/210$ the area of the regular island supports $23$ regular states.

The asymptotic state of the kicked rotor weakly coupled to a heat bath
is again determined from Eq.~\eqref{eq:RGS_rho_diag},
with the composite rates $R_{ij} = R_{ij}^{(1)} + R_{ij}^{(2)}$
from Eq.~\eqref{eq:R_ik_cyclic} appropriate for a cyclic system.
The resulting Floquet occupations $p_i$ are shown in
Fig.~\ref{fig:phSp_Emean_rho_Ereg_rho}(b) as functions of the cycle-averaged 
energies $\langle E_i \rangle$, with $E_0 = -\kappa/(2\pi)^2$.
The regular and chaotic states again are ordered with respect
to this quantity,
as indicated by the arrows in Fig.~\ref{fig:phSp_Emean_rho_Ereg_rho}(b).
The regular states have small values of $\langle E \rangle$,
since both kinetic and potential energies are minimal in the center of
the regular island, whereas the chaotic states have a stronger overlap with
regions of phase space with higher energies.
Similarly as for the driven oscillator, the regular occupations depend
monotonously on $\langle E \rangle$,
while the occupations of the chaotic states seem uncorrelated with the 
cycle-averaged energy $\langle E \rangle$ and form a plateau 
with only weak fluctuations around a mean value $\pch$.

\subsection{Rate matrix}\label{sec:Rates}

\begin{figure}[tb]
\includegraphics[width=8.5cm]{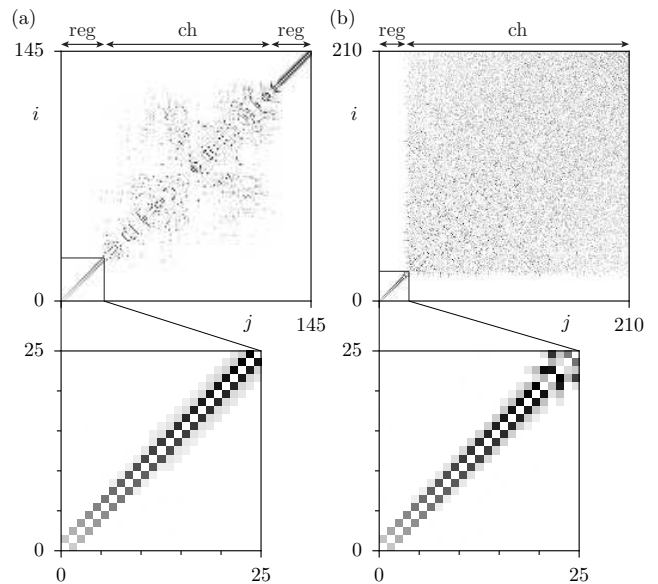}
\caption{
Density representation of the rate matrix $R_{ij}$,
with entries $i \neq j$ sorted by increasing $\langle E_i \rangle$ for
(a) the driven oscillator, Eq.~\eqref{eq:H_quart},
and (b) the kicked rotor, Eq.~\eqref{eq:H_kiRo},
with enlarged domain of regular island states,
revealing strong nearest-neighbor rates.
For parameters see Figs.~\ref{fig:phSp_Emean_rho_Ereg_rho_quart}
and~\ref{fig:phSp_Emean_rho_Ereg_rho}, respectively.
}
\label{fig:ratematrix}
\end{figure}

The Floquet rate matrix $R_{ij}$ determines the Floquet occupations
via Eq.~\eqref{eq:RGS_rho_diag}.
In Fig.~\ref{fig:ratematrix} we show $R_{ij}$ for both
the driven oscillator in Fig.~\ref{fig:phSp_Emean_rho_Ereg_rho_quart} 
(Fig.~\ref{fig:ratematrix}(a))
and the kicked rotor in Fig.~\ref{fig:phSp_Emean_rho_Ereg_rho}
(Fig.~\ref{fig:ratematrix}(b)).
Employing $\langle E_i \rangle$ as the ordering parameter for the entries $i,j$, 
the regular and chaotic parts are well-separated, revealing a distinct 
block structure of the matrix~\cite{BreHubPet2000}.
There are only few rates between the regular and the chaotic subspaces.
Similarly, the rates between the two different regular subspaces
in the case of the driven oscillator in Fig.~\ref{fig:ratematrix}(a) 
are practically zero.
Also by virtue of the chosen ordering, the regular domains feature a
band structure with particular dominance of the first off diagonals
(nearest-neighbor rates).
The rates in the subspace of the chaotic states on the contrary fluctuate
strongly.

One can thus observe a close relation between the structure of the rate matrix
$R_{ij}$ and the resulting set of occupations.
First, the almost independent behaviour of the occupations
of regular states and chaotic states owes to the relatively weak rates 
$R_{ij}$ connecting the corresponding subspaces.
Furthermore, the random character of the chaotic rate submatrix gives rise to
the equally random character of the set of chaotic Floquet occupations.
Note that for the kicked rotor
we find in the semiclassical limit $h \to 0$, that the mean value
$\pch$ decreases, since more and more regular states emerge.
Also the relative variance
$\overline{ \left( p_i - \pch \right)^2 } \big/ \pch^2$
of the chaotic occupations $p_i$ decreases in this limit
and we observe a universal scaling in the semiclassical limit
which can be analyzed with the help of a random-rate model~\cite{Wus2010}.
These aspects of the chaotic occupations are not explored in this paper,
instead the focus is set on the regular occupations.

\section{Regular States}\label{sec:RegularStates}

The observations in Figs.~\ref{fig:phSp_Emean_rho_Ereg_rho_quart}
and~\ref{fig:phSp_Emean_rho_Ereg_rho}
indicate that the asymptotic state of a time-periodic system in weak interaction
with a heat bath carries signatures of the classical phase-space structure.
The Floquet occupations of the regular and the chaotic states behave very
differently, e.g.\ as functions of the cycle-averaged energy $\langle E \rangle$.
In this section we focus on the asymptotic occupations of the regular states,
which we label by their quantum number $m$ starting with $m=0$ for the state
in the center of the island.
Figures~\ref{fig:phSp_Emean_rho_Ereg_rho_quart}(b)
and~\ref{fig:phSp_Emean_rho_Ereg_rho}(b)
suggest a roughly exponential dependence for the regular occupations $p_m$ as
functions of the cycle-averaged energies $\langle E_m\rangle$.
However, the regular occupations are different from
the Boltzmann weights $e^{-\beta \langle E_m \rangle}$
with the true inverse bath temperature $\beta$.
In fact, there is no physical reason for a coincidence with the Boltzmann
distribution when expressed in terms of the, qualitatively suitable
but arbitrary, energy measure $\langle E_m \rangle$ ~\cite{Koh2001}.
In the following sections we therefore make use of an alternative
energy measure for the regular states, the regular energy
$\Ereg_m$ (Sec.~\ref{sec:Ereg}), allowing us to consistently parametrize
the regular occupations as functions of $\Ereg_m$ (Sec.~\ref{sec:approx1_nN}).
Often, this functional dependence is approximately exponential
(Sec.~\ref{sec:approx2_betaeff}).
Examples are presented in Sec.~\ref{sec:examples}.

\subsection{Regular energy $\Ereg_m$}\label{sec:Ereg}

A time-periodic system is equivalent to an autonomous system
with the time as an additional coordinate, 
leading to the Hamiltonian $H_s'(x,p;t,p_t) = H_s(x,p;t) + p_t$ 
in the extended phase space,
which has periodic boundary conditions in $t$.
This allows the application of Einstein-Brillouin-Keller(EBK)-quantization rules 
for the regular tori
and the determination of semiclassical Floquet states on the quantizing tori
and their associated semiclassical 
quasienergies~\cite{BreHol1991,BenKorMirBen1992}.

We introduce the regular energies
\begin{equation}\label{eq:Ereg_m}
\Ereg_{m} :=
\hw \nu_m \left(m + \frac{1}{2}\right) - \left\langle L \right\rangle_m
+ \left\langle L \right\rangle_c
.
\end{equation}
Herein, $\nu_m$ is the winding number, 
i.e.\ the ratio of the winding frequency of a trajectory on the $m$th torus 
around the central orbit 
to the driving frequency $\omega$. 
Furthermore, $\left\langle L \right\rangle_m$
is the long-time average of the Lagrangian $L = p\dot x - H_s$
for an arbitrary trajectory on the $m$th torus.
For convenience we add the time-averaged Lagrange function
$\langle L \rangle_c$ of the central orbit of the island.
The regular energies of Eq.~\eqref{eq:Ereg_m} are related to the semiclassical 
quasienergies~\cite{BreHol1991,BenKorMirBen1992} by
\begin{equation}\label{eq:Eq_Ereg}
 \varepsilon_m = \Ereg_m - \langle L \rangle_c \mod \hw
,
\end{equation}
whereas there is no relation to the cycle-averaged energies 
$\langle E_m \rangle$.
The time-averaged Lagrangian $\langle L \rangle$
varies only slowly inside the island.
The winding number $\nu$ likewise varies slowly across the island.
To determine $\nu$ we have applied the frequency map
analysis~\cite{LasFroeCel1992} being based on a Fourier decomposition
of the quasiperiodic orbits within a stable regular island.
Note that due to nonlinear resonances and small chaotic layers within a regular
island the semiclassical quantization might require interpolations of the 
quantities $\nu_m$ and $\langle L \rangle_m$ or the introduction of a fictitious 
integrable system~\cite{BenKorMirBen1992, BohTomUll1993_BaeKetLoeSch2008}.

\begin{figure}[tb]
\includegraphics[width=8.5cm]{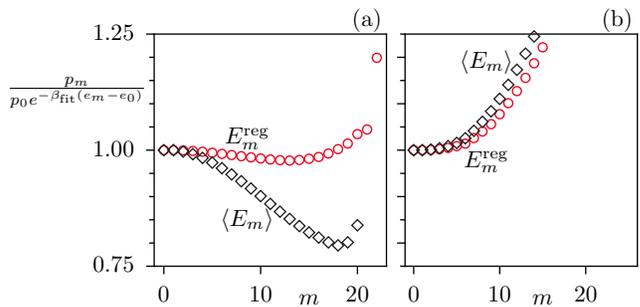}
\caption{
(Color online) 
Ratio between the occupations $p_m$ and an exponential fit
$p_0 e^{-\betafit (e_m-e_0)}$
for (a) the kicked rotor
and (b) the continuously driven oscillator
using the regular energies $e_m = \Ereg_m$ (red circles)
and the cycle-averaged energies $e_m = \langle E_m \rangle$ (black diamonds).
For parameters see Figs.~\ref{fig:phSp_Emean_rho_Ereg_rho}
and~\ref{fig:phSp_Emean_rho_Ereg_rho_quart}, respectively.
}
\label{fig:Ereg_Emean_devfit}
\end{figure}

Figure~\ref{fig:phSp_Emean_rho_Ereg_rho}(c) shows
the occupations $p_m$ of the regular states of the kicked rotor
as functions of regular energies $\Ereg_m$.
The functional dependence of the occupations $p_m$ is close to exponential, but
also different from the Boltzmann weights $e^{-\beta \Ereg_m}$
with the true bath temperature $1/\beta$.
However, the assumption of an exponential dependence of $p_m$ vs.\ $\Ereg_m$ is
fulfilled far better than vs.\ $\langle E_m \rangle$.
This is demonstrated in Fig.~\ref{fig:Ereg_Emean_devfit}, where the
ratio between the occupations $p_m$ and the respective exponential fit
$p_0 e^{-\betafit (e_m-e_0)}$ is shown
for $e_m$ being the regular energy $\Ereg_m$ (red circles) and
the cycle-averaged energy $\langle E_m \rangle$ (black diamonds).
The fit involves the parameter $\betafit := (\log p_1 - \log p_0)/(e_0-e_1)$.
For the kicked rotor, Fig.~\ref{fig:Ereg_Emean_devfit}(a),
the considered ratio for $e_m = \Ereg_m$ is close to $1$ for the majority of 
regular states, whereas the ratio for $e_m = \langle E_m \rangle$ 
systematically deviates from $1$ already for smaller values of $m$.
This indicates that the exponential scaling is far better fulfilled
by using the regular energies $\Ereg_m$.

Likewise, Fig.~\ref{fig:Ereg_Emean_devfit}(b) shows the same ratio 
for the regular states of the central island in the
driven oscillator, Eq.~\eqref{eq:H_quart}.
The regular energies are again determined according to the
above semiclassical quantization, where the frequency map analysis is applied
to the solutions of the classical equations of motion,
evaluated in the Poincar\'{e} section.
Here, the quality of the fit with respect to $\Ereg_m$
is only marginally better as the the fit with respect to $\langle E_m \rangle$,
see Fig.~\ref{fig:Ereg_Emean_devfit}(b).
We have evidence, that the existence of next-nearest-neighbor rates
is responsible for this.
Note that for other examples of continuously driven systems
we typically find a better quality of the fit with respect to $\Ereg_m$,
similar to the situation in Fig.~\ref{fig:Ereg_Emean_devfit}(a).

\subsection{Restriction to nearest-neighbor rates $R_{m,m\pm 1}$}%
\label{sec:approx1_nN}

In this section, the ratio of the rates $R_{m,m+1}$ and $R_{m+1,m}$ between
two neighboring regular states $m$ and $m+1$ is analyzed for kicked systems.
With the help of a detailed balance condition the occupations $p_m$ can be
related to the winding numbers of the regular tori.

In the lower part of Fig.~\ref{fig:ratematrix}(b) the rate matrix
$R_{ij} = R_{ij}^{(1)} + R_{ij}^{(2)}$, Eq.~\eqref{eq:R_ik_cyclic},
for the regular subspace of the kicked rotor is shown.
We remind the reader that the indices
are ordered by increasing $\langle E \rangle$, coinciding with
the natural order of growing quantum number $m$.
Figure~\ref{fig:ratematrix}(b) illustrates that the nearest-neighbor rates
$R_{m,m \pm 1}$ are dominant among the regular states.
These nearest-neighbor rates are mainly contributed by the rates $R_{ij}^{(1)}$
originating from the coupling operator $A^{(1)} = \sin(2\pi x)/(2\pi)$,
whereas the rates $R_{m,m \pm 2}$ between next-nearest neighboring states
are mainly due to the contribution $R_{ij}^{(2)}$ of the
coupling operator $A^{(2)} = \cos(2\pi x)/(2\pi)$.
In the following analytical considerations
we will neglect the contribution of $R_{ij}^{(2)}$ and in addition
approximate the coupling operator $A^{(1)} = \sin(2\pi x)/(2\pi)$
inside the regular island at $x=0.5$ by the linear coupling operator $A=-x$.
Using the resulting rate matrix in Eq.~\eqref{eq:RGS_rho_diag},
we observe almost the same regular occupations as
in Fig.~\ref{fig:phSp_Emean_rho_Ereg_rho}(b),
which in first approximation differ from the latter
only by a tiny $m$-independent factor.
Only the occupations of the chaotic states are strongly affected by the
different coupling scheme, as for them the discontinuity of
the coupling operator $x$ at the border of the unit cell is not negligible.

In the rate matrix due to the linear coupling operator, $A=-x$,
the nearest-neighbors dominate strongly,
and the next-nearest-neighbor rates $R_{m,m \pm 2}$
of regular states $m$ and $m \pm 2$ are zero for symmetry reasons.
We will neglect higher-order rates in the following analysis.
Thus, the total rate balance among the regular states
can be reduced to the detailed balance condition
\begin{equation}\label{eq:detbal_approx}
\frac{p_{m+1}}{p_{m}} = \frac{R_{m,m+1}}{R_{m+1,m}}
\end{equation}
between two neighboring regular states $m$ and $m+1$.

Assuming Eq.~\eqref{eq:detbal_approx} and using the definition of 
the rates in Eq.~\eqref{eq:R_ik} the occupation ratio 
between neighboring regular states becomes
\begin{eqnarray}
\lefteqn{
\frac{p_{m+1}}{p_{m}}
} \nonumber\\
&=&
\frac{
\sum_K \left| A_{m,m+1}(K) \right|^2
g\left( \varepsilon_{m+1}-\varepsilon_{m} - K\hw \right)
}{
\sum_K \left| A_{m+1,m}(K) \right|^2
g\left( \varepsilon_{m}-\varepsilon_{m+1} - K\hw \right)
}  \\
&=&
\label{eq:R_ij_rel}
\frac{
\sum_K \left| x_m(K) \right|^2
g\left( \zeta_m + K\hw \right)
}{
\sum_K \left| x_m(K) \right|^2
g\left( \zeta_m + K\hw \right)
e^{\beta (\zeta_m + K\hw)}
}
,
\end{eqnarray}
where the properties $A_{m,m+1}(K) = A_{m+1,m}^{*}(-K)$ 
and $g(E) = g(-E) e^{-\beta E}$ have been used,
and where we introduced the short-hand notations
\begin{equation}
 x_m := A_{m+1,m}
\end{equation}
for the (regular) nearest-neighbor matrix elements of the operator $A=-x$ and
\begin{equation}
\zeta_m := \varepsilon_{m+1} - \varepsilon_m
\end{equation}
in the arguments of the correlation function $g(E)$.

If there were just a single Fourier component $x_m(K^{*})$,
which is approximately the case for a weakly driven system,
then the occupation ratio would simplify to
$p_{m+1}/p_{m} = e^{-\beta (\zeta_m + K^{*}\hw)}$, resulting in 
Boltzmann-like occupations.
In general, however, several components $K$ have to be considered.

Multiplying numerator and denominator of the fraction in Eq.~\eqref{eq:R_ij_rel}
with $\left| x_m(0) \right|^{-2} g\left( \zeta_m \right)^{-1}$,
\begin{equation}\label{eq:R_ijji_kicked}
\frac{p_{m+1}}{p_{m}} = 
 \frac{
 \sum_K \frac{ \left| x_m(K) \right|^2 }{ \left| x_m(0) \right|^2 }
 \frac{ g\left( \zeta_m + K\hw \right)
 }{ g\left( \zeta_m \right) }
 }{
 \sum_K \frac{ \left| x_m(K) \right|^2 }{ \left| x_m(0) \right|^2 }
\frac{
  g\left( \zeta_m + K\hw \right)
 }{ g\left( \zeta_m \right) }
 e^{\beta (\zeta_m + K\hw)}
 }
,
\end{equation}
we introduce ratios of the matrix elements and of the correlation functions.
The ratio of the correlation functions reads,
using their definition in Sec.~\ref{sec:FloquetMarkov},
\begin{eqnarray}\label{eq:ratio_g}
\frac{g(\zeta_m + K \hw)}{g(\zeta_m)}
 &=&
\left( 1 + \frac{K\hw}{\zeta_m}\right)
\frac{ e^{\beta \zeta_m} - 1
}{
e^{\beta \left(\zeta_m + K\hw\right)} - 1} \\
&& \cdot
\exp \left(
\frac{\left|\zeta_m\right| -\left|\zeta_m + K\hw \right|
}{ \hbar \omega_c }
\right) \nonumber
.
\end{eqnarray}
The last factor in Eq.~\eqref{eq:ratio_g} is close to $1$ and will be
omitted in the following,
such that the $\omega_c$-dependence is neglected. 
This is possible
since we are interested here in the case $\omega_c \gg \omega$
and we use the fact that only small integers $K$ contribute significantly 
to the sums in Eq.~\eqref{eq:R_ijji_kicked}.

For the required ratio of the matrix elements
we approximate the evolution of the coupling matrix elements $x_m(t)$
for $0 \leq t \leq \tau$ by
\begin{equation}\label{eq:x_ijt_2}
x_m(t) \approx x_m(t=0) e^{-\iexp \zeta_m t/\heff}
\left[
1 + \frac{t}{\tau} \left( e^{\iexp \zeta_m \tau/\heff} - 1 \right)
\right]
,
\end{equation}
using the factorization of the time evolution operator for kicked systems
and the approximate commutation relations with the operator $x$
on the two-torus,
$\left[x, e^{-\iexp V(x) \tau /\heff}\right] \approx 0$
and
$\left[x, e^{-\iexp T(p) t/\heff} \right] \approx t T'(p) e^{-\iexp T(p) t/\heff}$.
These commutation relations, which are exact in the infinite Hilbert space,
apply here in very good approximation to the regular states, as these are
almost independent of the periodic boundary conditions on the two-torus.
The coupling matrix elements $x_m(t)$ are time-periodic and 
have the Fourier components
\begin{equation}\label{eq:x_ijK}
x_m(K) = \frac{x_m(t=0)}{2 \pi^2}
\left(\frac{\zeta_m}{\hw} + K\right)^{-2}
\left(1-\cos\left( 2\pi \frac{\zeta_m}{\hw}  \right)\right)
,
\end{equation}
whose ratio simplifies to
\begin{equation}\label{eq:ratio_x2}
 \frac{\left| x_m(K) \right|^2}{\left| x_m(0) \right|^2}
= \left( 1 + \frac{K\hw}{\zeta_m} \right)^{-4}
\;.
\end{equation}

Finally, inserting Eqs.~\eqref{eq:ratio_g} and~\eqref{eq:ratio_x2}
into the occupation ratio of Eq.~\eqref{eq:R_ijji_kicked} yields
\begin{equation}\label{eq:R_ijji_kicked_2}
\frac{p_{m+1}}{p_{m}} =
F\left( \frac{\zeta_m}{\hw}, \beta\hw \right)
\end{equation}
with the function
\begin{eqnarray}
F\left(z,b\right) :=
\frac{
\sum_K \left(K + z\right)^{-3} \left( e^{(K+z)b} - 1 \right)^{-1}
}{
\sum_K \left(K - z\right)^{-3} \left( e^{(K-z)b} - 1 \right)^{-1}
}
\;.
\end{eqnarray}
It is invariant under an integer shift of the first argument, 
$F(z + K_0, b) = F(z,b)$, with $K_0 \in \mathbb{Z}$
\cite{comment1}.
We choose the shift $K_0$, such that 
\begin{equation}\label{eq:zeta_Ereg}
\varepsilon_{m+1}-\varepsilon_m + K_0\hw = \Ereg_{m+1}-\Ereg_m 
\end{equation}
is fulfilled, which is possible according to Eq.~\eqref{eq:Eq_Ereg}.
This allows us to replace $\zeta_m$ in Eq.~\eqref{eq:R_ijji_kicked_2}
with the regular energy spacing $\Ereg_{m+1,m} := \Ereg_{m+1} - \Ereg_{m}$,
leading to  
\begin{equation}
\frac{p_{m+1}}{p_{m}} = F\left( \frac{\Ereg_{m+1,m}}{\hw}, \beta \hw \right)
.
\end{equation}
Based on Eq.~\eqref{eq:Ereg_m} we approximate this energy difference by
the winding number
\begin{equation}
\Ereg_{m+1,m} \simeq \hw \nu_m
,
\end{equation}
which is exact for a harmonic oscillator-like island
with $m$-independent winding number $\nu_m$ and $\langle L \rangle_m$
and is a reasonable approximation even for more generic islands.
The occupation ratio then becomes a function of the winding number,
\begin{equation}\label{eq:dist_preg0}
\frac{p_{m+1}}{p_{m}} = F\left( \nu_m, \beta \hw \right)
.
\end{equation}
With that an analytical prediction for the occupation of the regular
states of a kicked system is found,
valid under the assumption of dominant nearest-neighbor rates $R_{m,m \pm 1}$.
It is a function of the winding number of the $m$th quantizing torus
and of the parameters temperature and driving frequency.

\subsection{Assumption of constant winding number $\nu$}%
\label{sec:approx2_betaeff}

\begin{figure}[tb]
\includegraphics[width=5.2cm]{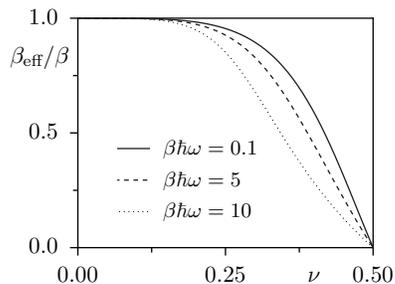}
\caption{
Inverse effective temperature $\betaeff/\beta$
according to Eq.~\eqref{eq:betaeff} vs.\ winding number $\nu$
for $\omega_c/\omega \gg 1$ and three different temperatures.
}
\label{fig:nu_betaeff}
\end{figure}

The function $F$ in Eq.~\eqref{eq:dist_preg0} becomes independent of $m$
if the winding number $\nu$ is constant throughout the regular island.
It is then appropriate to introduce an effective temperature $1/\betaeff$ by
\begin{equation}\label{eq:betaeff}
\betaeff := -\frac{\log F\left(\nu, \beta \hw\right)}{\hw \nu}
.
\end{equation}
With this new parameter the occupation ratios are expressed in a form
analogous to the Boltzmann weights,
\begin{equation}\label{eq:dist_preg}
p_{m+1}/p_m = e^{-\betaeff \Ereg_{m+1,m}}
.
\end{equation}

The ratio $\betaeff/\beta$ is shown in Fig.~\ref{fig:nu_betaeff}.
Its value is smaller than $1$ and it is symmetric in $\nu$.
For $\nu \to 0$, where the kicked system approaches
its static limit, the true bath temperature $1/\beta$ is retained.
A substantial deviation from the true bath temperature, $\betaeff/\beta \ll 1$,
takes place around $\nu \approx 0.5$.

For generic islands with non-constant winding number,
where the regular energy spacings $\Ereg_{m+1,m} \simeq \hw \nu_m$
are $m$-dependent,
the exponential scaling, Eq.~\eqref{eq:dist_preg}, is still approximately
valid if $\betaeff$ varies only moderately.
Figure~\ref{fig:nu_betaeff} indicates that this is fulfilled especially good
for values $\nu \lesssim 0.2$.
But even beyond this interval we observe a good agreement of
Eq.~\eqref{eq:dist_preg} with the regular occupations.
This breaks down for islands with winding numbers varying close to $\nu=0.5$,
where $\betaeff$ is particularly sensitive to variations of $\nu$.
As $\betaeff$ develops a pronounced $m$-dependence there,
the approximation~\eqref{eq:dist_preg} is no longer adequate
and the general equation~\eqref{eq:dist_preg0} has to be used instead,
as demonstrated in Fig.~\ref{fig:phSp_Emean_rho_Ereg_rho_dev} below.

\subsection{Examples}\label{sec:examples}

Figure~\ref{fig:phSp_Emean_rho_Ereg_rho}(c) shows the regular Floquet
occupations of the kicked rotor, Eq.~\eqref{eq:H_kiRo}, vs.\
the regular energies $\Ereg_m$ and demonstrates excellent agreement
with the above predicted exponential weights $e^{-\betaeff \Ereg_m}$ (red solid line)
for almost all regular states.
The regular energies $\Ereg_m$ are here slightly $m$-dependent, as the
winding number decreases in the island monotonously
from $\nu_0 = 0.32$ for the first ($m=0$) regular state
to $\nu_{22} = 0.28$ ($m=22$) for the outermost regular state.
Deviations from the exponential distribution
occur for the outermost regular states, $m \geq 20$, only.
These have a stronger weight outside the regular island
and are thus coupled stronger to  the chaotic states.
The non-negligible rates between these regular states and the chaotic states,
see Fig.~\ref{fig:ratematrix},
enforce a gradual adaptation between the outermost regular occupations
and the occupation level of the chaotic states.
Besides, Fig.~\ref{fig:phSp_Emean_rho_Ereg_rho}(c) shows the large
discrepancy of the Boltzmann weights
$e^{-\beta \Ereg_m}$ with the true bath temperature $1/\beta$.

\begin{figure}[tb]
\includegraphics[width=8.5cm]{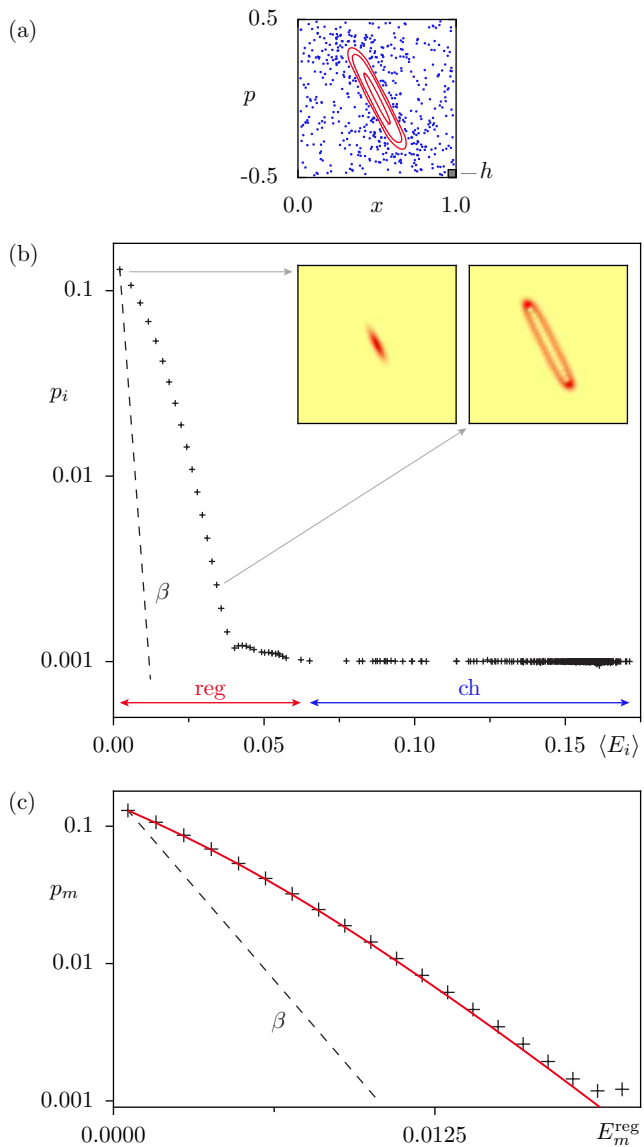}
\caption{
(Color online) 
(a)--(c) Analogous to Fig.~\ref{fig:phSp_Emean_rho_Ereg_rho} for the
kicked rotor with $\kappa = 3.9$.
The insets in (b) show Husimi representations of two regular states.
In (c) the occupations of the innermost $20$ regular states, which are not yet
affected by phase-space structures at the border of the island, are well
described by the analytical prediction (red solid line) 
of Eq.~\eqref{eq:dist_preg0}.
The parameters are $h = 1/400$ and $\beta = 500$.
}
\label{fig:phSp_Emean_rho_Ereg_rho_dev}
\end{figure}

The exponential distribution, Eq.~\eqref{eq:dist_preg},
requires a constant or moderately varying winding number inside the island.
If however $\nu$ varies close to $\nu=0.5$,
where $\betaeff$ is particularly sensitive to variations of $\nu$,
see Fig.~\ref{fig:nu_betaeff}, the exponential distribution is no longer
adequate.
For the kicked rotor this is the case for $\kappa \to 4$,
where the central periodic orbit bifurcates and the regular island
hence splits into two islands.
As an example, Fig.~\ref{fig:phSp_Emean_rho_Ereg_rho_dev}
shows the occupations for the kicked rotor for $\kappa = 3.9$,
with $\nu_0 = 0.44$ and $\nu_{19} = 0.38$.
The regular occupations in Fig.~\ref{fig:phSp_Emean_rho_Ereg_rho_dev}(c)
are well described by the general prediction, Eq.~\eqref{eq:dist_preg0}
(red line), but clearly deviate from the exponential approximation,
Eq.~\eqref{eq:dist_preg}.

The derivation of the analytical occupation ratios~\eqref{eq:dist_preg0}
and~\eqref{eq:dist_preg}
is based on assumptions which are justified for kicked systems only.
For continuously driven systems an analogous prediction remains a future
challenge.
Figure~\ref{fig:phSp_Emean_rho_Ereg_rho_quart}(b) demonstrates that
a significant and systematic deviation of the occupations from
the Boltzmann result is observed also for continuously driven systems.

For another continuously driven system, the driven particle in a box,
the regular occupations are almost identical to the
Boltzmann weights $e^{-\beta \langle E_m \rangle}$
with the true temperature $1/\beta$~\cite{BreHubPet2000}.
However, in this particular example
the regular region is almost identical to the undriven system,
leading to Boltzmann weights for the regular states by the following reasoning:
the regular states of the driven box potential, which emerge from the highly
excited eigenstates of the undriven box, still strongly resemble the latter and
change only slightly during the driving period $\tau = 2\pi/\omega$. 
By that,
one dominant Fourier contribution $K^{*}$ in the coupling matrix elements
is singled out, $x_m(K) \approx 0$ for $K \neq K^{*}$.
In this situation the occupation ratio~\eqref{eq:R_ij_rel}
simplifies to $p_{m+1}/p_{m} = e^{-\beta (\zeta_m + K^{*}\hw)}$
and the detailed balance among the regular states, Eq.~\eqref{eq:detbal_approx},
is fulfilled accurately.
Since at the same time their cycle-averaged energies
are close to the eigenenergies in the undriven potential,
the occupations are close to the Boltzmann weights.
Tiny deviations for the lowest regular states close to the chaotic region
are visible in Fig.~1 of Ref.~\cite{BreHubPet2000}, which we attribute to rates
occurring between the regular and the chaotic Floquet states.

As substantiated in this section, systematic and much stronger deviations from
the Boltzmann behavior can occur in generic situations,
especially in situations characteristic for strong driving, where
the phase-space structure and the Floquet states
are strongly perturbed compared to the original time-independent system.
The strong driving allows us to study the Floquet occupations
far from the thermodynamic equilibrium situation,
encountered in the time-independent system, 
and at the same time to have dominant
regular structures present in the classical phase space.
These host, a sufficiently semiclassical regime presumed,
a series of regular states,
which under the condition of pronounced nearest-neighbor rates
and only small rates to the subspace of the chaotic states
are occupied with weights given by Eq.~\eqref{eq:dist_preg0}.
For constant or slowly varying winding number $\nu$ the occupations even
simplify to the exponential weights, Eq.~\eqref{eq:dist_preg},
with the effective temperature of Eq.~\eqref{eq:betaeff}.

A generic modification of the analytical predictions of this section,
takes place as a consequence of avoided crossings.
We will go back to this point in Sec.~\ref{sec:AC}.

\section{Implications of additional classical phase-space structures}%
\label{sec:AddStructs}

The set of  Floquet states in the examples of the last section are dominated
by regular states in large regular islands and chaotic states.
Apart from these, other types of Floquet states can exist, depending on the
structures in the classical phase space and the size of the effective
Planck constant $h$. The following section gives an overview of the
fingerprints of such additional types of Floquet states on the distribution
of the Floquet occupations $p_i$.

\subsection{Nonlinear resonance chains}

\begin{figure}[tb]
\includegraphics[width=8.5cm]{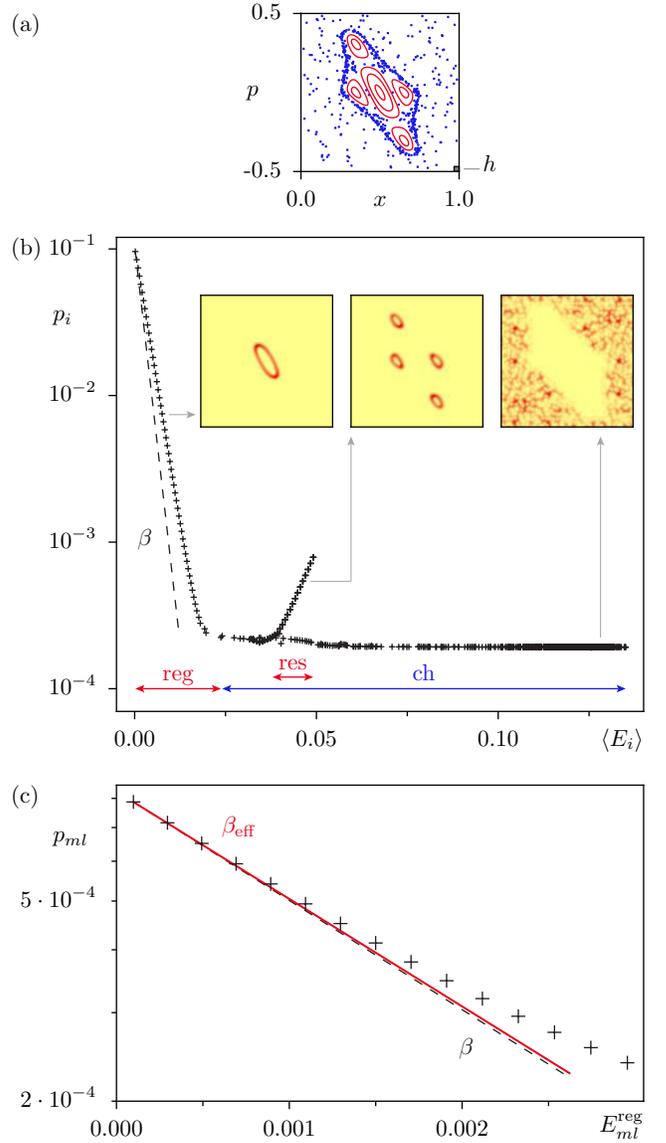}
\caption{
(Color online) 
(a) and (b) Analogous to Fig.~\ref{fig:phSp_Emean_rho_Ereg_rho} for the
kicked rotor with $\kappa = 2.35$
in presence of a $4$:$1$-resonance chain.
The insets in (b) show Husimi representations of a regular state from the main
island, a regular resonance state, and a chaotic state.
(c) Floquet occupations $p_{ml}$ of the regular resonance states
vs.\ regular energies $\Ereg_{ml}$,
the Boltzmann-like prediction Eq.~\eqref{eq:dist_preg} with
$\betaeff\approx 0.98\,\beta$ (red solid line)
compared to the inverse bath temperature $\beta$ (dashed line).
The parameters are $h = 1/1000$ and $\beta = 500$.
}
\label{fig:phSp_Emean_rho_Ereg_rho_res4}
\end{figure}

Apart from the islands centered at stable elliptic fixed points of period 1,
there are nonlinear $r$:$s$-resonances consisting of $r$ regular islands
around stable periodic orbits of period $r$,
see Fig.~\ref{fig:phSp_Emean_rho_Ereg_rho_res4}(a).
A trajectory on such a resonance chain passes from an island
to the $s$th next island
and returns after $r$ periods to the island, where it initially started.
Considering the $r$--fold iterated map instead of the map itself,
the trajectory always remains on one and the same island.

The semiclassical quantization is done with respect to this $r$--fold map of
period $r \tau$~\cite{MirKor1994}.
To each principal quantum number $m$ there exist $r$ regular Floquet states
$|\psi_{ml}\rangle$ of different quantum numbers $l=0,\ldots,r-1$ with
equidistant quasienergy spacing $\hw/r$.
We refer to these states as regular resonance states.
Each of them has equal weights in each of the dynamically connected resonance
islands, but with different phases.

Similarly as in Eq.~\eqref{eq:Ereg_m} we derive from the
semiclassical quasienergies the corresponding regular energies
\begin{equation}\label{eq:Ereg_ml}
\Ereg_{ml} =  \hw \frac{\nu_m^{(r)}}{r} \left(m+\frac{1}{2}\right)
- \left\langle L \right\rangle_{m}
+ \left\langle L \right\rangle_c
,
\end{equation}
which are independent of the quantum number $l$.
The winding number $\nu_m^{(r)}$ refers to the $r$--fold iterated map.

Figure~\ref{fig:phSp_Emean_rho_Ereg_rho_res4}(b) shows the Floquet occupations
$p_i$ vs.\ the cycle-averaged energy $\langle E_i\rangle$ for the kicked rotor
with $\kappa = 2.35$, where the phase space features in addition to the main
regular island a $4$:$1$-resonance around the periodic orbit of period $4$,
see Fig.~\ref{fig:phSp_Emean_rho_Ereg_rho_res4}(a).
The entire resonance chain hosts $4\cdot15$ regular resonance states 
for $h=1/1000$.
The Floquet occupations of both the regular states of the central island and
the chaotic states resemble those of Fig.~\ref{fig:phSp_Emean_rho_Ereg_rho}(b).
In addition, one finds a branch belonging to the regular resonance states.
Interestingly, it has a positive slope stemming from the fact that the
cycle-averaged energies $\langle E_{ml} \rangle$ of the regular resonance states 
$|\psi_{ml}\rangle$ 
decrease with increasing quantum number $m$,
in contrast to the regular states of the central island.
This is due to the asymmetry of the resonance tori around their respective
island center in phase space.
This is another clear evidence, that the cycle-averaged energy does not serve
as a suitable measure to quantify the regular occupations by exponential
weights in analogy to the Boltzmann distribution.

The $r$ regular resonance states $|\psi_{ml}\rangle$ of fixed quantum number $m$
have almost the same cycle-averaged energy $\langle E_{ml} \rangle$.
As long as the coupling to the heat bath does not disturb the equivalence
of the resonance islands,
the occupations $p_{ml}$ of the $r$ regular resonance states of fixed principal
quantum number $m$ are independent of the quantum number $l$.
In Fig.~\ref{fig:phSp_Emean_rho_Ereg_rho_res4}(b)
the corresponding four branches of the occupations $p_{ml}$
therefore lie almost on top of each other
and cannot be distinguished on the scale of the figure.
Small deviations from this degeneracy exist only for the outermost
regular resonance states.
These can be attributed to the occurrence of avoided crossings,
which  break the degeneracy of
$\varepsilon_{ml} \mod (\hw/r)$ for $l=0,\ldots,r-1$ and fixed $m$,
as well as the degeneracy of $\langle E_{ml} \rangle$ and $p_{ml}$.

Now we want to explain that the occupations $p_{ml}$ of the regular resonance
states are likewise distributed as $p_{ml} \sim e^{-\betaeff \Ereg_{ml}}$,
according to the exponential weights~\eqref{eq:dist_preg}
with the effective temperature $1/\betaeff$ of Eq.~\eqref{eq:betaeff}.
We note that Eq.~\eqref{eq:ratio_g} 
for the ratio of the correlation functions
and Eq.~\eqref{eq:ratio_x2} for the ratio of the coupling matrix elements
apply without restriction also to the regular resonance states.
The assumed detailed balance relation, Eq.~\eqref{eq:detbal_approx}, however,
is no longer adapted to the structure of the rate matrix since here,
in addition to the nearest-neighbor rates $R_{(ml)(m \pm 1,l)}$,
also `internal' rates exist, i.e.\ rates in the subspace of
the $r$ equivalent regular resonance states $l=0,\ldots,r-1$
with fixed quantum number $m$.
Nonetheless, the total rate balance approximately decouples for each principal
quantum number $m$ into the $r$ balance relations for $l=0,\ldots,r-1$
\begin{equation}\label{eq:detbal_approx_res_2}
\frac{ p_{m+1,l} }{p_{ml}} \approx
\frac{ R_{(ml)(m+1,l)} }{ R_{(m+1,l)(ml)} }
.
\end{equation}
They have the same structure as Eq.~\eqref{eq:detbal_approx}
and turn out to be approximately $l$-independent, leading to approximately
$l$-independent occupations $p_{ml}$.
In Eq.~\eqref{eq:detbal_approx_res_2}
the tiny rates $R_{(ml)(m'l')}$ with $m \neq m'$ and $l \neq l'$
are neglected and one can show that the contribution
$\sum_{l'} \left( R_{(ml)(ml')} - R_{(ml')(ml)} \right)$
vanishes as a consequence of the equidistant quasienergy spacing
for the $r$ regular resonance states of the same $m$~\cite{Wus2010}.

The decoupling into the $r$ equivalent balance
relations~\eqref{eq:detbal_approx_res_2} finally allows us to approximate
the occupations $p_{ml}$ by the exponential weights $e^{-\betaeff \Ereg_{ml}}$
of Eq.~\eqref{eq:dist_preg} with the effective temperature $1/\betaeff$ of
Eq.~\eqref{eq:betaeff}.
Figure~\ref{fig:phSp_Emean_rho_Ereg_rho_res4}(c)
shows the occupations $p_{ml}$ of the regular resonance states
vs.\ the regular energies $\Ereg_{ml}$.
Even on the magnified scale of this subfigure, 
compared to Fig.~\ref{fig:phSp_Emean_rho_Ereg_rho_res4}(b),
the tiny differences of the occupations $p_{ml}$ with different
quantum numbers $l$ are not visible.
The effective temperature $1/\betaeff$ is nearly indistinguishable
from the actual temperature $1/\beta$, because the winding number
$\nu_{m=0}^{(r)} /r = 0.79 /4 \approx 0.2$
of the resonance islands is small and yields a
value of $\betaeff/\beta$ very close to $1$, compare Fig.~\ref{fig:nu_betaeff}.
Note that it differs, although weakly, from $\betaeff \approx 0.93$
of the main island.
The parameter $\betaeff$ is the same for each of the four
independent occupation branches $p_{ml}$.

The phase space of a generic time-periodic system contains
a hierarchy of nonlinear resonance chains and islands of all scales.
If $h$ is sufficiently small, one has Floquet states on these islands and
we expect that the entire set of Floquet occupations becomes increasingly 
structured by the branches originating from each nonlinear resonance.

\subsection{Beach states}

\begin{figure}[tb]
\includegraphics[width=8.5cm]{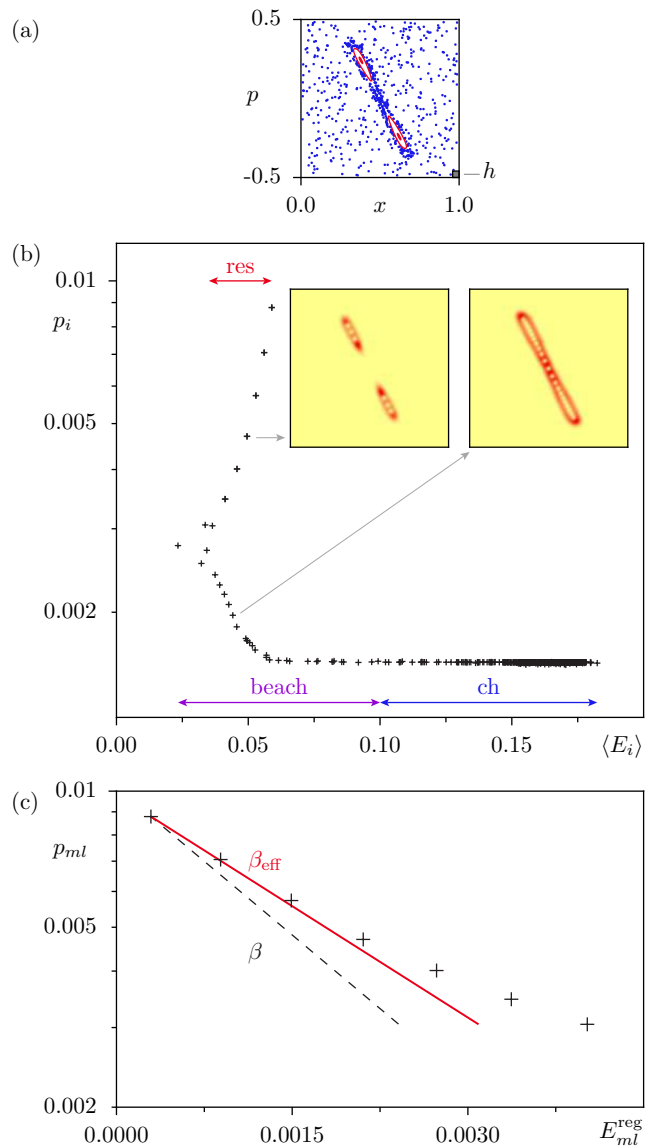}
\caption{
(Color online) 
(a) and (b) Analogous to Fig.~\ref{fig:phSp_Emean_rho_Ereg_rho} for the
kicked rotor with $\kappa=4.415$ in the presence of a $2$:$1$-resonance
surrounded by a series of strong partial barriers.
The insets in (b) show Husimi representations of a regular resonance state
($m=3$) and a beach state.
(c) Floquet occupations $p_{ml}$ of the regular resonance states
vs.\ regular energies $\Ereg_{ml}$,
the Boltzmann-like prediction Eq.~\eqref{eq:dist_preg} with
$\betaeff\approx 0.76\,\beta$ (red solid line)
compared to the inverse bath temperature $\beta$ (dashed line).
The parameters are $h = 1/600$ and $\beta = 500$.
}
\label{fig:phSp_Emean_rho_Ereg_rho_res2}
\end{figure}

The transition between regular phase-space regions and
the chaotic sea is usually not
sharp, but shaped by a multitude of small island chains
and cantori, the fractal remains of broken Kolmogorov-Arnol'd-Moser tori.
These additional phase-space structures can strongly inhibit the classical
flux of trajectories toward and away from the regular island
and, depending on the size of $h$,
can give rise to the formation of quantum beach states, a term
introduced in Ref.~\cite{FriDor1998}.
These reside on the transition layer around the regular islands
and have little overlap with the remaining chaotic sea.
Typically, beach states have very similar appearance and properties like
the regular states of the adjacent island.
Due to the proximity they partly even allow a quantization
similar to the EBK-quantization rules~\cite{FriDor1998, BohTomUll1990}.

At $\kappa = 4$ the central island of the kicked rotor bifurcates into
a resonance around a stable periodic orbit of period $2$.
It is accompanied by a series of partial barriers
with a reduced classical flux toward and away from the islands.
This is indicated for $\kappa = 4.415$
in the stroboscopic Poincar\'{e} section
of Fig.~\ref{fig:phSp_Emean_rho_Ereg_rho_res2}(a)
by the relatively high density of the chaotic orbit
in the vicinity of the island.
Figure~\ref{fig:phSp_Emean_rho_Ereg_rho_res2}(b) shows the Floquet
occupations $p_i$ vs.\ the cycle-averaged energy $\langle E_i \rangle$.
The highest occupations belong to the regular states of the resonance.
The occupations of the beach states form a separate, nearly monotonous set in 
the transition region between the occupations of the regular resonance states
and the chaotic states.
This is a consequence of the structure in the coupling matrix $R_{ij}$,
where typically the nearest-neighbor rates dominate, 
similarly as for the regular states.

Furthermore, the regular occupations $p_{ml}$ of the regular resonance states
are shown vs.\ $\Ereg_{ml}$ in Fig.~\ref{fig:phSp_Emean_rho_Ereg_rho_res2}(c).
In this example the winding number $\nu_{m=0}^{(r)} /r = 0.71 /2 \approx 0.35$
in the resonance islands
yields a stronger deviation between $\beta$ and $\betaeff$,
with $\betaeff/\beta \approx 0.76$, than in the example presented in 
Fig.~\ref{fig:phSp_Emean_rho_Ereg_rho_res4}.

\subsection{Hierarchical states}

As mentioned above, in the vicinity of regular islands typically many
partial barriers with a limited classical flux toward and away from the
island can be found, e.g.\ in the form of cantori or based on
stable and unstable manifolds~\cite{MaKMeiPer1992_Mei1992}.
Depending on the ratio of $h$ to the classical flux,
partial barriers can prevent Floquet states from spreading
over the entire chaotic domain, apart from tunneling tails.
If the phase-space area enclosed by the island and the partial barrier
exceeds $h$, these states locally resemble chaotic states.
For decreasing values of $h$ they resolve and occupy the hierarchy
of the classical phase space better and better and are therefore
called hierarchical states~\cite{KetHufSteWei2000}.
The existence of these states does not contradict the semiclassical
eigenfunction hypothesis~\cite{Per1973_Ber1977_Vor1979},
as their fraction vanishes with $\mathcal{O}(h^{\alpha})$
in the semiclassical limit.
We apply an overlap criterion to determine, whether a Floquet state is 
hierarchical or not: it is identified as a hierarchical state,
if it is not a regular state but  comparably strongly localized,
such that its Husimi weight $\iint_\Omega \diff x \diff p H_\psi(x,p)$
within a large chaotic phase-space area $\Omega$ away from the regular
island falls below $70\%$,
compared to a state that is uniformly spread over the entire phase space.

\begin{figure}
\includegraphics[width=8.5cm]{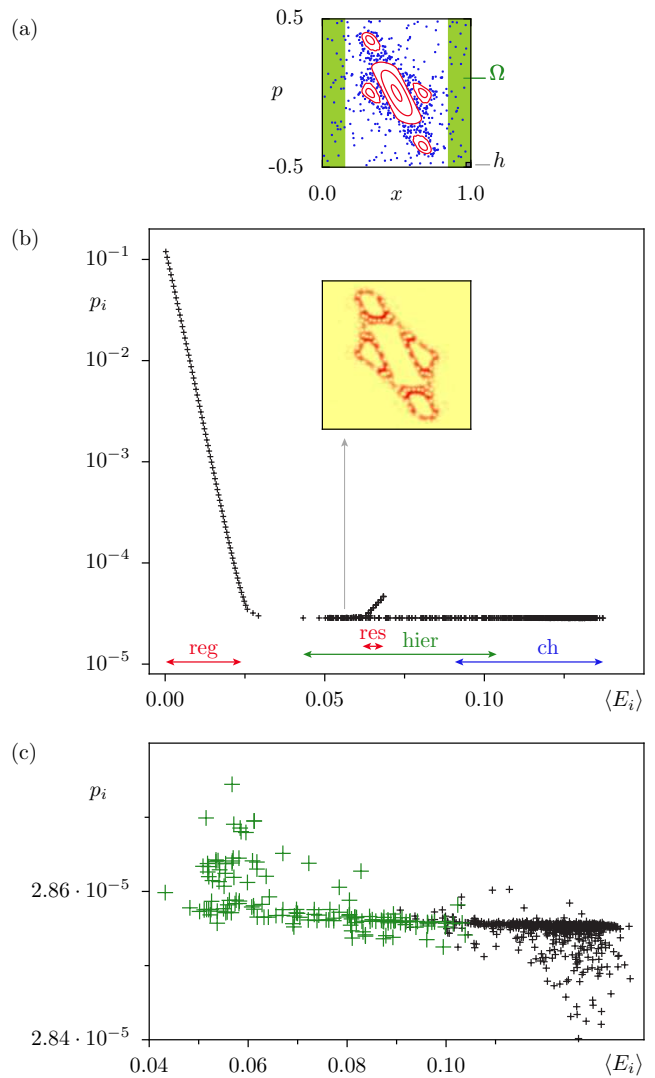}
\caption{
(Color online) 
(a) and (b) Analogous to Fig.~\ref{fig:phSp_Emean_rho_Ereg_rho} for the
kicked rotor with $\kappa = 2.5$
in the presence of a $4$:$1$-resonance surrounded by a partial barrier.
The inset in (b) is the Husimi representation of a hierarchical state.
The branch with positive slope belongs to the regular resonance states
like in Fig.~\ref{fig:phSp_Emean_rho_Ereg_rho_res4}(b).
(c) Magnification of (b) with emphasized data points of the hierarchical states
(large green crosses), which are determined by the overlap criterion
from the shaded phase-space area $\Omega$ in (a).
The parameters are $h = 1/1000$ and $\beta = 500$.
}
\label{fig:phSp_Emean_rho_Ereg_rho_hier}
\end{figure}

Figure~\ref{fig:phSp_Emean_rho_Ereg_rho_hier}(a) shows the Poincar\'{e} section
and Fig.~\ref{fig:phSp_Emean_rho_Ereg_rho_hier}(b) shows the Floquet occupations
for the kicked rotor with $\kappa=2.5$, where the fraction of hierarchical
states is comparatively high~\cite{KetHufSteWei2000}.
In Fig.~\ref{fig:phSp_Emean_rho_Ereg_rho_hier}(c),
the occupations of the hierarchical states are emphasized.
The figure indicates that their occupations are distributed analogously
to the chaotic states which explore the entire chaotic phase-space region.
Again, the fluctuation pattern of the occupations $p_i$
has its origin in the randomly fluctuating rates $R_{ij}$ in the subspace
of the hierarchical states, as is the case for the chaotic states.

To conclude this section,
the occupation characteristics of the beach states and the hierarchical states
again confirm the influence of the classical phase-space structure not only
on the spectrum and on the Floquet states, but eventually also on the
Floquet occupations and hence on the asymptotic state of the system.

Note that in the above examples, Figs.~\ref{fig:phSp_Emean_rho_Ereg_rho_res2}
and~\ref{fig:phSp_Emean_rho_Ereg_rho_hier}, either of the two types
is predominant, but still representatives of the other are present.
In general, hierarchical and beach states coexist.
For example, a few of the states of intermediate cycle-averaged energy 
$\langle E \rangle$
that are indicated in Fig.~\ref{fig:phSp_Emean_rho_Ereg_rho_hier}
as hierarchical by the above overlap criterion had rather to be classified as
beach states or as states with scarring behavior,
i.e.\ localized on hyperbolic fixed points or on a family
of parabolic fixed points.

\section{Avoided crossings}\label{sec:AC}

Since the spectrum of Floquet systems is restricted to a finite
interval $0 \leq \varepsilon<\hw$,
a multitude of avoided level crossings typically emerges
under the variation of a parameter and
gives rise to the hybridization of the involved Floquet states.
In the case of an infinite dimensional Hilbert space
the quasienergy spectrum is dense and there is no longer an adiabatic limit,
i.e.\ any tiny parameter variation will hybridize
infinitely many Floquet states in a complex way~\cite{HonKetKoh1997}.
However, as shown in Ref.~\cite{HonKetKoh2009},
the asymptotic density operator $\rho$ is not
affected by a small avoided crossing, provided that it
is smaller than a specific effective coupling strength to the heat bath.
Thus, the interaction with the heat bath resolves
the difficulties of the dense quasienergy spectrum.
In this section we focus on the opposite limit, where a single isolated avoided
crossing strongly influences the entire set of Floquet occupations.

\begin{figure}[tb]
\includegraphics[width=8.5cm]{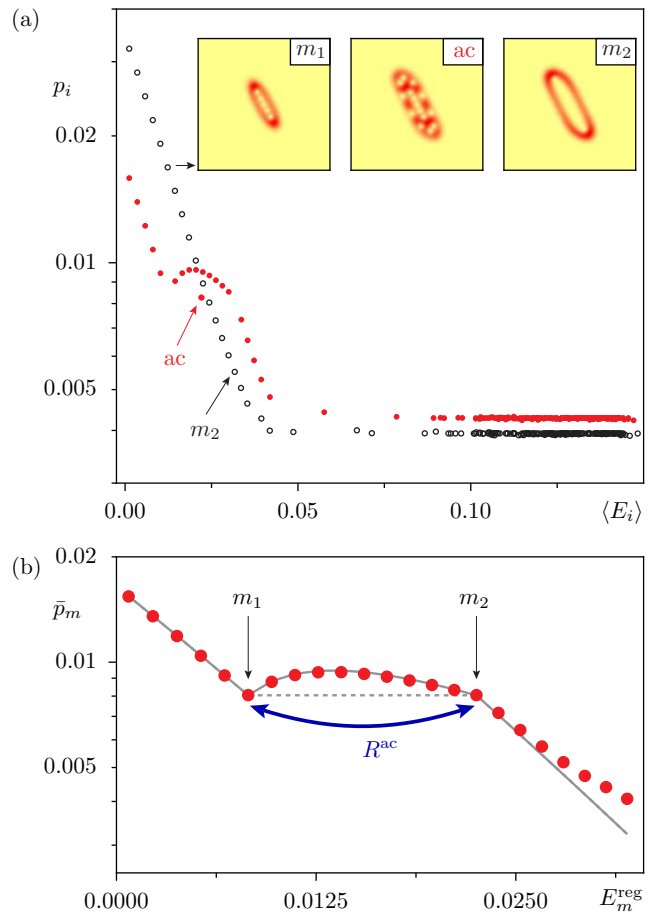}
\caption{
(Color online) 
Influence of an avoided crossing between the regular states $m_1=5$ and $m_2=15$
on Floquet occupations.
(a) Floquet occupations $p_i$ vs.\ cycle-averaged energies $\langle E_i \rangle$
for $\kappa=\kappa_1=2.85$ (black circles) and
$\kappa=\kappa_2=2.857175$ (red dots) close to the 
center of the avoided crossing.
The insets show the Husimi representations of states $m_1$, $m_2$ at $\kappa_1$
and of a corresponding hybridized state at $\kappa_2$ (`ac').
(b) Occupations $\bar p_m$ of regular states vs.\ regular energies $\Ereg_m$
at $\kappa = \kappa_2$ (red dots)
and comparison to the analytical solution~\eqref{eq:rho_model2}
of the rate model (solid gray line) and the model with $R_{m,m+1}=R_{0,1}$
from~\cite{HonKetKoh2009} (dashed gray line).
The arrow indicates the effective rate $\Rac$ between states $m_1$ and $m_2$
according to Eq.~\eqref{eq:Rac}.
Note that $\bar p_{m_1}$ and $\bar p_{m_2}$ are measured in the diabatic basis,
in contrast to $p_i$ in (a).
The parameters are $h = 1/210$ and $\beta=100$.
}
\label{fig:Emean_rho_Ereg_rho_AC1}
\end{figure}

Figure~\ref{fig:Emean_rho_Ereg_rho_AC1} presents a typical
example of the kicked rotor.
In Fig.~\ref{fig:Emean_rho_Ereg_rho_AC1}(a)
the Floquet occupations $p_i$ are shown vs.\ $\langle E_i \rangle$
for two values of the kick strength near $\kappa = 2.9$, very close to the
parameter realization in Fig.~\ref{fig:phSp_Emean_rho_Ereg_rho}.
The difference of these $\kappa$-values is sufficiently small,
such that the classical phase space
and almost all regular states vary only marginally.
For two of the regular Floquet states, which we will denote as states
$a$ and $b$, and which are initially identical
to the semiclassical modes with the quantum numbers $m_1=5$ and $m_2=15$,
this is however not the case. Under the variation of $\kappa$ they undergo
an avoided crossing at $\kappa \approx 2.857$, where they hybridize.

The tiny $\kappa$-variation strongly affects the Floquet occupations,
and most prominently all the regular occupations.
Away from the avoided crossing ($\kappa_1=2.85635$),
the regular occupations monotonously decrease with $\langle E_i \rangle$,
similar as in Fig.~\ref{fig:phSp_Emean_rho_Ereg_rho}(b).
When approaching the center of the avoided crossing ($\kappa_2=2.857175$)
the monotonous behaviour is locally disturbed:
the states $a$ and $b$, as a consequence of their hybridization,
have shifted mean energies $\langle E_{a} \rangle$ and $\langle E_{b} \rangle$
as well as modified occupations $p_{a} \approx p_{b}$.
In Fig.~\ref{fig:Emean_rho_Ereg_rho_AC1}(a) the data points of
the states $a$ and $b$ at $\kappa_2$ (marked as `ac')
are therefore found indistinguishable on top of each other.
Beyond that, also the occupations $p_m$ of all regular states with
quantum numbers $m$ from the interval $[m_1,m_2]$ change severely.
They are close to the value of occupation $p_a \approx p_b$ of the
hybridized states.
In contrast, their mean energies $\langle E_m \rangle$ do not change notably
under the tiny $\kappa$-variation, like those of the semiclassical
modes $m_1$ and $m_2$.
The relative occupations $p_m / p_{m+1}$ among the regular states
with quantum numbers outside the range $[m_1,m_2]$
are also not affected. Only the absolute values of their occupations $p_m$
are shifted due to the normalization $\sum_i p_i = 1$.
The latter is also the origin of a shift of the chaotic occupation 
plateau $\pch$.

\begin{figure}[tb]
\includegraphics[width=8.5cm]{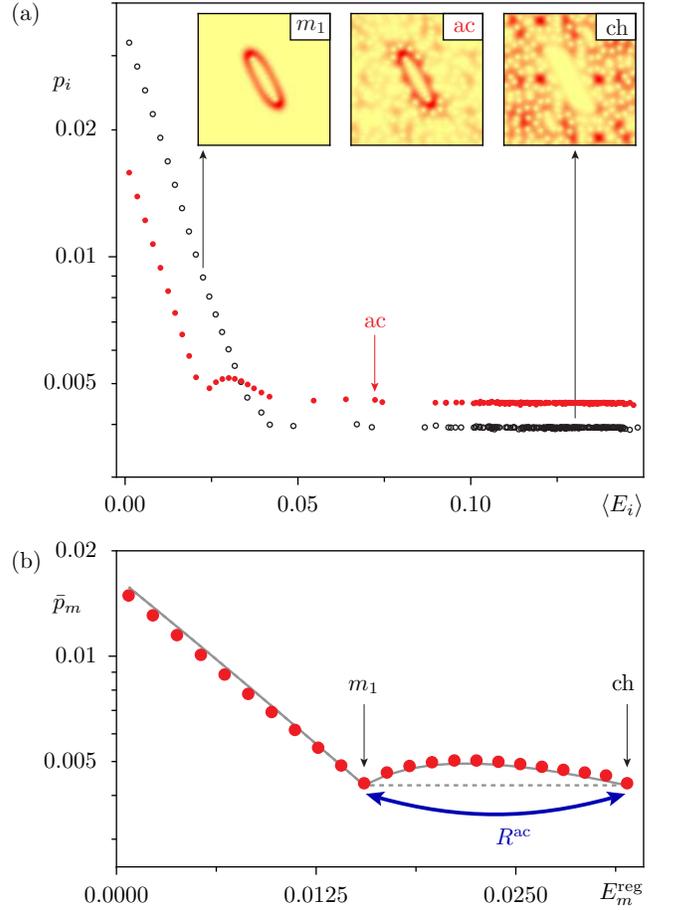}
\caption{
(Color online) 
Influence of an avoided crossing between the regular state $m_1=10$
and a chaotic state on the Floquet occupations,
analogous to Fig.~\ref{fig:Emean_rho_Ereg_rho_AC1}
for $\kappa=\kappa_1=2.856400$ (black circles) 
and $\kappa=\kappa_2=2.856897$ (red dots)
close to the center of the avoided crossing.
Note in (b), that we have simulated the avoided crossing of $m_1$
with a chaotic state in the rate model, Eq.~\eqref{eq:rho_model2},
by an avoided crossing between $m_1$ and $m=22$, the last regular state.
The parameters are $h = 1/210$ and $\beta=100$.
}
\label{fig:Emean_rho_Ereg_rho_AC2}
\end{figure}

This example demonstrates that the presence of avoided crossings can change
the entire character of the occupation distribution.
To explain this impact the authors of Ref.~\cite{HonKetKoh2009}
introduced an effective rate equation,
\begin{equation}\label{eq:RGS_rho_AC}
0 = - \bar p_i \sum_j \bar R_{ij} + \sum_j \bar p_j \bar R_{ji}
.
\end{equation}
which refers to a representation in the local diabatic basis
of the avoided crossing, denoted by an overbar.
In the diabatic basis the states $a$ and $b$
are replaced with states that remain invariant at the
avoided crossing, i.e.\ the semiclassical modes $m_1$ and $m_2$
in the case of an avoided crossing of two regular states.
In Eq.~\eqref{eq:RGS_rho_AC} the typically negligible rates
$\bar R_{m_1 m_2}$, $\bar R_{m_2 m_1}$ are replaced with 
a new effective rate~\cite{HonKetKoh2009}
\begin{equation}\label{eq:Rac}
\Rac := \frac{\Gamma}{\left(\hbar \Gamma / \Delta\right)^2 + 4 d^2}
,
\end{equation}
which acts between the states $m_1$ and $m_2$.
The gap size $\Delta$,
i.e.~the minimal quasienergy splitting $|\varepsilon_a-\varepsilon_b|$
of states $a$ and $b$
and the dimensionless distance from its center
$d = (\bar \varepsilon_{m_1} - \bar \varepsilon_{m_2})/\Delta$
are characteristic properties of the avoided crossing.
Unlike the rates $\bar R_{ij}$, which are nearly constant in the vicinity
of the avoided crossing, the additional rate $\Rac$ changes dramatically.
The composite rate
$\Gamma = \sum_{j} \left( \bar R_{m_1 j} + \bar R_{m_2 j} \right)
- (2\pi/\hbar) \sum_K \bar A_{m_1 m_1}(K) \bar A_{m_2 m_2}(K) g(-K\hw)$
plays the role of an effective coupling strength and the characteristic
parameter $\hbar\Gamma/\Delta$ determines, whether $\Rac$ can become
dominant at the center of the avoided crossing $d=0$.
In the examples of Figs.~\ref{fig:Emean_rho_Ereg_rho_AC1}
and~\ref{fig:Emean_rho_Ereg_rho_AC2}
the condition $\hbar\Gamma/\Delta \ll 1$ is fulfilled and the rate
$\Rac$ consequently dominates around $d \approx 0$ with respect to other rates
in Eq.~\eqref{eq:RGS_rho_AC}.
Note that this requires the system-bath coupling strength to be sufficiently
small, a limit which has been already presumed in Eq.~\eqref{eq:RGS_rho_diag}.
The dominance of $\Rac$ is responsible for
the local disruption of the formerly exponential behavior
of the regular occupations:
as $\Rac$ exceeds all other rates, it induces occupation equality
of the states $m_1$ and $m_2$.
The relative occupations of the regular states below $m_1$ and above $m_2$
still follow the approximate detailed balance, Eq.~\eqref{eq:detbal_approx},
with rates acting predominantly
between nearest neighbors.
Finally, the other regular states,  $m_1 < m < m_2$,
although also still having dominant nearest-neighbor rates,
no longer have exponentially scaling occupations, since the additional
rate $\Rac$ between $m_1$ and $m_2$ breaks the detailed balance.
These conditions explain the observed signature of the avoided
crossing in the occupation characteristics and are substantiated by a
simplified rate model introduced in the following section.

A second example, with an avoided crossing between the regular state $m_1=10$
and a chaotic state, is presented in Fig.~\ref{fig:Emean_rho_Ereg_rho_AC2}(a).
Here, the occupation of the state $m_1$ is forced down to the
chaotic occupation level in the vicinity of the avoided crossing (red dots).
Moreover, the entire subset of occupations $p_m$ with $m > m_1$ is disturbed
from the original exponential scaling (black circles).
With a similar reasoning as above,
this behavior is explained by the additional rate $\Rac$ and a simplified
rate model, see below.

These consequences of avoided crossings also explain why the
transition between the occupations of regular and chaotic states
appears so 'smoothly', see e.g.~Figs.~\ref{fig:phSp_Emean_rho_Ereg_rho_quart}
and~\ref{fig:phSp_Emean_rho_Ereg_rho}.
As the outer regular states or the beach states most likely
undergo avoided crossings with chaotic states,
the occupation probabilities normally
feature a transition to the plateau $\pch$ of the chaotic occupations.

We emphasize that the observed implication of an
avoided level crossing is a remarkable effect bound
to the non-equilibrium character of the driven system,
with possible applications, e.g.\
bath-induced switching in a driven double well potential~\cite{KetWus2009}.
In a time-independent system on the contrary,
an avoided crossing entails only a local shift in the occupations
of the two involved states and leaves the Boltzmann distribution of the entire
set of occupations unchanged.

\subsection{Simplified rate model}\label{sec:AC_model}

To account for the local modifications of the occupations $p_m$,
which are entailed by an avoided crossing,
we apply a simplified, analytically solvable model for the set of
rate equations~\eqref{eq:RGS_rho_AC}.
It is based on the nearest-neighbor rate model
in Ref.~\cite{HonKetKoh2009}.
It is restricted to the chain of regular states and,
at first in the absence of an avoided crossing,
assumes that each of them is coupled only to its directly neighboring states.
Under this circumstance the detailed balance is fulfilled.
The primary parameter of the model is the rate ratio of the first two states,
$g := R_{0,1}/R_{1,0}$.
We approximate the rate ratio $R_{m,m+1}/R_{m+1,m}$ as independent of $m$,
i.e.\ $R_{m,m+1}/R_{m+1,m} \equiv g$ for all $m$.
According to Eqs.~\eqref{eq:dist_preg0} and~\eqref{eq:detbal_approx}
the approximation is especially suited for regular islands with a constant
winding number $\nu$.

The rates themselves however differ from state to state.
In contrast to Ref.~\cite{HonKetKoh2009}, where $R_{m,m+1}/R_{0,1}$
is presumed as constant, we make use of the approximation
\begin{equation}\label{eq:R_mmp1_R_01}
\frac{R_{m,m+1}}{R_{0,1}} = m+1
.
\end{equation}
This relation holds exactly for the states of a harmonic oscillator
with a linear coupling to the heat bath.
The coupling matrix elements fulfill $A_{m,m+1} = \sqrt{m+1}\: A_{0,1}$,
a property that is easily shown with the help
of the associated algebra of ladder operators.
Analogously one can prove relation~\eqref{eq:R_mmp1_R_01} for the
regular states of a time-periodic system if the regular tori
are elliptic, coinciding with those of a harmonic oscillator,
and assume $m$-independent winding numbers.
Beyond the application to elliptic islands of constant winding number
we presume Eq.~\eqref{eq:R_mmp1_R_01} for arbitrary islands,
which seems to be in general a good approximation.

To account for an avoided crossing between the states $m_1$ and $m_2$
the model introduces the additional rates $R_{m_1 m_2} = R_{m_2 m_1} = \Rac$
from Eq.~\eqref{eq:Rac}.
By the action of $\Rac$ the flux
$F_{m,m-1} := p_m R_{m,m-1} -  p_{m-1} R_{m-1,m}$
becomes non-zero for all $m_1 < m \leq m_2$.
The rate equations, Eq.~\eqref{eq:RGS_rho_AC}, translate to the flux equations
\begin{eqnarray}\label{eq:flux}
0 &=& F_{1,0} \\
0 &=& F_{m+1,m} - F_{m,m-1}  \qquad {\scriptstyle m \neq 0, m_1, m_2}\nonumber\\
 \left( p_{m_1} - p_{m_2} \right) \Rac &=& F_{m_1+1,m_1} - F_{m_1,m_1-1}
\nonumber \\
-\left( p_{m_1} - p_{m_2} \right) \Rac &=& F_{m_2+1,m_2} - F_{m_2,m_2-1}
\nonumber
\end{eqnarray}
with the solution
$F_{m,m-1} = F := \left( p_{m_1} - p_{m_2} \right) \Rac$ for $m_1 < m \leq m_2$
and $F_{m,m-1} = 0$ otherwise.
Eventually, the occupations assume the values
\begin{equation}\label{eq:rho_model2}
p_m = \left\{ \begin{array}{l l}
p_0 \hspace{2mm} g^m
& {\scriptstyle m \leq m_1}\\
p_{m_1} \!\left[ \left(1 - g^{m_2-m_1}\right)\frac{r_m}{1+r_{m_2}} + g^{m-m_1} \right]
& {\scriptstyle m_1 < m \leq m_2} \\
p_{m_2} g^{m-m_2}
& {\scriptstyle m_2 \leq m}
\end{array} \right.
\end{equation}
with
$r_m := \Rac \sum_{k=1}^{m-m_1} \left( g^{m-m_1-k} / R_{m_1+k,m_1+k-1} \right)$.
For $\Rac \gg R_{1,0}$ the parameter $r_{m_2}$ diverges and
$p_{m_2}/p_{m_1}
= \left(1 - g^{m_2-m_1}\right) \left[r_{m_2}/(1+r_{m_2})\right] + g^{m_2-m_1}$
approaches $1$.
Note that the model solution~\eqref{eq:rho_model2}
relies on the nearest-neighbor coupling and on the
$m$-independence of $R_{m,m+1}/R_{m+1,m}$.
The ratio $R_{m,m+1}/R_{0,1}$ of Eq.~\eqref{eq:R_mmp1_R_01} leads to
$r_m = \left(\Rac/R_{1,0}\right) \sum_{k=1}^{m-m_1} \left[ g^{m-m_1-k}/(m_1+k)\right]$.

In Figs.~\ref{fig:Emean_rho_Ereg_rho_AC1}(b)
and~\ref{fig:Emean_rho_Ereg_rho_AC2}(b) we apply the simplified rate
model to regular states for a kick strength close to the center
of the avoided crossing, and compare its solution, Eq.~\eqref{eq:rho_model2},
to the occupations $\bar p_m$ from the solution of
the rate equations~\eqref{eq:RGS_rho_diag}.
The regular energies $\Ereg$ are not well defined for the hybridizing
states of the avoided crossing, but are well defined for the respective
diabatic states $m_1$ and $m_2$.
That is why the occupations in these subfigures are represented
in the diabatic basis by means of the orthogonal transformation
$\bar p_{m_1} = \alpha^2 p_{a} + \beta^2 p_{b}$
and $\bar p_{m_2} = \beta^2 p_{a} + \alpha^2 p_{b}$
with $\alpha^2 = \left(1+d/\sqrt{1+d^2}\right)/2$ and $\alpha^2 + \beta^2 = 1$.
The comparison indicates that the rate model based on
assumption~\eqref{eq:R_mmp1_R_01} reproduces the local disturbance of the
exponential scaling for the states $m_1 < m < m_2$ very accurately.
In contrast, the simpler assumption~\cite{HonKetKoh2009} of
$m$-independent rates, $R_{m,m+1}/R_{0,1} = 1$,
does not reproduce the $m$-dependence of the occupations between $m_1$ and $m_2$
(dashed line).

For the example in Fig.~\ref{fig:Emean_rho_Ereg_rho_AC2}(b)
the rate model seems at first sight not applicable,
since it does not account for chaotic states.
We therefore apply the model as if the avoided crossing were between the
regular state $m_1=10$ and the outermost regular state, $m=22$, and still
obtain a good agreement between the observed $p_m$ and the model solutions.
This is possible, as the occupation of the regular state $m=22$
differs only weakly from the plateau of chaotic occupations.

\FloatBarrier
\section{Summary}\label{sec:Summary}

A core question of statistical mechanics is the characterization of the
asymptotic state approached by a quantum system when it interacts with a
thermal reservoir.
In the familiar equilibrium thermodynamics of time-independent systems in the
weak-coupling limit it is answered by the canonical distribution,
where the eigenstates of the isolated quantum system
are occupied with the statistical weights $p_i \sim e^{-\beta E_i}$.
In a time-periodic quantum system, where an external field permanently pumps
energy into the system and prevents its relaxation to equilibrium,
this is in general an intricate question, which cannot be answered by
deduction from the time-independent case.
Here, the asymptotic state under a weak coupling
to the thermal reservoir becomes time-periodic
and is best characterized by the time-independent occupations $p_i$ of the
Floquet states.
We demonstrate, that the Floquet occupations can be classified
according to the semiclassical character of the Floquet states.
The occupations of the chaotic Floquet states fluctuate weakly around
a mean value $\pch$~\cite{BreHubPet2000}.
The regular Floquet states on the contrary acquire probabilities that are
roughly exponentially distributed.
The validity of this observation is also confirmed by the occupation
characteristics of other types of Floquet states, which still reflects
their regular or chaotic nature:
beach states, which are very similar to the regular states and situated close,
but outside the regular island, they form a correlated set of occupations,
which is qualitatively comparable to the regular occupations.
In contrast, the occupations of hierarchical states,
which have the properties of chaotic states, but live
in a restricted region of the chaotic phase space,
are distributed analogously to the chaotic states.

In contrast to previous studies of a driven particle in a
box~\cite{BreHubPet2000}, where the regular states carry occupations close to
the Boltzmann weights, we observe that in general
the regular occupations can considerably deviate from the Boltzmann result.
This observation is possible as we focus on time-periodic systems
where the classical phase space and the Floquet states
are strongly perturbed compared to the originally time-independent system.
In kicked systems the occupations of the regular states can be well
approximated by a function $F(\nu_m, \beta\hw)$
depending on the classical winding numbers $\nu_m$ of the regular tori,
and the parameters temperature $1/\beta$ and driving frequency $\omega$.
For a constant or sufficiently moderately varying winding number 
within the regular island the distribution of the regular occupations can even
be described by weights of the Boltzmann type, $p_m \sim e^{-\betaeff \Ereg_m}$,
depending on the regular energies $\Ereg_m$.
The effective temperature $1/\betaeff$
is evaluated as a function of the winding number in the regular island.
As the driven system is not in an equilibrium state,
the proper definition of a non-equilibrium temperature
is a subtle problem~\cite{Rug1997_MorRon1999_CasJou2003}.
The critical question for future investigations is,
whether the quantity $\betaeff$ is accessible by a measurement.

A situation, where purely classical information is
no longer sufficient to account for the observed occupations,
is present at avoided crossings, which are ubiquitous in the quasienergy
spectra of Floquet systems.
Avoided crossings involving a regular state
give rise to strong changes in the set of Floquet occupations.
We give an intuitive explanation of this effect,
based on the additional rate $\Rac$ from Ref.~\cite{HonKetKoh2009},
and introduce a simplified rate model whose analytical solutions describe
the numerical data accurately.

In conclusion, with the presented characterizations of the Floquet occupations
we demonstrate that ubiquitous signatures of the classical dynamics 
are reflected in the asymptotic density matrix of the open quantum system.
In this way it is feasible to draw an intuitive picture of the asymptotic state,
shedding light on the statistical mechanics of time-periodic quantum systems.

\section*{Acknowledgements}

We acknowledge helpful discussions with S.~Fishman, D.~Hone, W.~Kohn, T.~Kottos,
and S.~L\"ock.
We thank the DFG for support within the Forschergruppe 760 
``Scattering Systems with Complex Dynamics''
and R.K.~thanks the Kavli Institute for Theoretical Physics at UCSB 
(NSF Grant No. PHY05-51164).

\end{document}